\documentclass[11pt]{article}

\usepackage[margin=.75in]{geometry}

\usepackage{color}

\definecolor{indigo}{RGB}{0,0,120}
\usepackage[colorlinks=true, linkcolor=indigo, citecolor=blue, urlcolor=indigo, backref=page]{hyperref}

\def\tr{\;{\rm tr}\;}

\def\pr{\prime}

\def\fl{\noindent}

\newcommand{\diag}{\: {\rm diag}}

\newcommand{\tl}[1]{\tilde{#1}}

\newcommand{\dd}[2]{\frac {\partial #1}{\partial #2}}

\newcommand{\pdr}{\partial}

\newcommand{\grad}{{\bm \nabla}}

\newcommand{\beq}{\begin{equation}}
	\newcommand{\eeq}{\end{equation}}
\newcommand{\beqs}{\begin{eqnarray}}
	\newcommand{\eeqs}{\end{eqnarray}}

\newcommand{\half}{\frac{1}{2}}
\newcommand{\ov}[1]{\frac{1}{#1}}

\newcommand{\fdaho}{\rm fd\text{-}aho}

\def\al{\alpha}		\def\g{\gamma} 		 
\def\del{\delta}	\def\D{\Delta}		\def\eps{\epsilon} 
			\def\ka{\kappa} 
\def\la{\lambda}		\def\sig{\sigma}		
		\def\tht{\theta}	 
		\def\om{\omega}		  
\def\z{\zeta}		
\def\Ups{\Upsilon}

\usepackage{titlesec}
\titleformat{\section}{\normalsize\bfseries}{\thesection}{1em}{}
\titleformat{\subsection}{\small\bfseries}{\thesubsection}{1em}{}
\titleformat{\subsubsection}{\small\bfseries}{\thesubsubsection}{1em}{}

\usepackage{amsmath,amssymb} % for AMS symbols and macros

\usepackage{graphicx}

\usepackage{calligra} % for script and cursive characters like script r
\DeclareMathAlphabet{\mathcalligra}{T1}{calligra}{m}{n}
\DeclareFontShape{T1}{calligra}{m}{n}{<->s*[2.2]callig15}{}

\newcount\colveccount  % column vec \colvec{nrows}{a11 & a12}{a21 & a22}
\newcommand*\colvec[1]{\global\colveccount#1  \begin{pmatrix} \colvecnext} \def\colvecnext#1{#1 \global\advance\colveccount-1
		\ifnum\colveccount>0 \\ \expandafter\colvecnext
		\else \end{pmatrix} \fi}

\newenvironment{smmat}
{\left(\begin{smallmatrix}}
	{\end{smallmatrix}\right)}

\usepackage[skip=3pt,labelfont=normal,labelsep=colon]{caption}
\usepackage[skip=3pt]{subcaption} %skip decides the spacing between subfigures and subcaptions, default=6pt.%If you want (a) to be in bold face: labelfont=bf. %default labelsep=space.
\usepackage{wrapfig}

\usepackage{cancel}
\usepackage[normalem]{ulem}
\usepackage{bm}

\newcommand{\tlk}{{\tl k}}

\newcommand{\bma}{{\bm a}}
\newcommand{\bmb}{{\bm b}}

\newcommand{\bme}{{\bm e}}
\newcommand{\bmf}{{\bm f}}

\newcommand{\bmp}{{\bm p}}

\newcommand{\bmr}{{\bm r}}

\newcommand{\bmA}{{\bm A}}
\newcommand{\bmB}{{\bm B}}
\newcommand{\bmE}{{\bm E}}

\newcommand{\calO}{{\cal O}}

\begin{document}

\title{\Large Lax pairs and $r$-matrices for some two-dimensional isotropic oscillators}

\author{{\sc Govind S. Krishnaswami$^{a}$, Garima Rajpoot$^{b}$ and Srividya N. Vaidya$^{c}$}
\\ \small
$^{a}$Physics Department, Chennai Mathematical Institute,  SIPCOT IT Park, Siruseri 603103, India\\ \small
$^{b}$Nuclear Physics Division, Bhabha Atomic Research Centre, Mumbai 400085, India\\
\small
$^{c}$Department of Physics, Indian Institute of Science Education and Research Tirupati, Yerpedu 517619, India \\
\small
Email: {\tt govind@cmi.ac.in, garimar@barc.gov.in, srividya$\_$20221143@students.iisertirupati.ac.in}}

\date{July 23, 2026}

\maketitle

%---------------------
\vspace{-1cm}
%---------------------

\abstract{\small This paper concerns Lax pairs for circularly symmetric harmonic, Fock-Darwin-type and quartic anharmonic oscillators in two dimensions. Although the 2d isotropic harmonic oscillator is bi-Hamiltonian, its recursion operator does not lead to a Lax pair, nor do we obtain such a pair by taking a limit of the harmonic Calogero model. On the other hand, we show that this superintegrable harmonic oscillator admits a $4 \times 4$ block-form Lax pair with spectral parameter giving two conserved mode energies in involution and a corresponding dynamical $r$-matrix. Interestingly, we also find $2 \times 2$ Lax pairs with spectral parameter that give all three independent conserved quantities satisfying a nonabelian Poisson algebra, thereby providing a simple example of a Lax pair whose conserved quantities are not all in involution. Next, we construct a family of $su(2)$ Lax pairs and $r$-matrices for the quadratic+quartic isotropic anharmonic oscillator. This is then extended to an isotropic oscillator with a rotational energy, which may be viewed as the Fock-Darwin oscillator with a quartic potential. With a change of variables, these Lax pairs and $r$-matrices also apply to the Rajeev-Ranken model, although its noncanonical Poisson structure is distinct from that of the anharmonic oscillator.}

% abstract: 183 words, 1216 characters

\scriptsize

\tableofcontents

\normalsize

%-----------------
\section{Introduction and Summary}
\label{s:intro}
%-----------------

Two-dimensional oscillators such as the harmonic, Fock-Darwin \cite{fock-1928,darwin-1930} and anharmonic oscillators arise in many ways: modeling a particle on a plane trapped in a potential, a planar charged particle subject to an axial magnetic and radial electric field, mechanical reduction of a field theory \cite{rajeev-ranken,gsk-trv-rr-mod-anharm-osc}, etc. They have practical applications, such as to quantum dots \cite{quantum-dot-anharm-osc} and optical traps \cite{2d-harm-trap-bose-gas} but have also served as simple yet insightful contexts in which to explore physical and mathematical phenomena and develop techniques \cite{drigho-kuru-negro-nieto-fock-darwin,sinha-ghosh-bagchi-anisotropic-osc-2023}. Among oscillators, the integrable ones form an interesting subclass where many properties can be studied through exact methods. In this paper, we focus on Lax pairs and associated structures for classical isotropic 2d oscillators, which are integrable since they admit two commuting conserved quantities: total energy and angular momentum. Lax pairs \cite{peter-lax} are foundational to integrable systems: a pair of matrices $A$ and $B$ whose entries depend on the dynamical variables in such a way that the Lax equations $\dot A = [B,A]$ are equivalent to the equations of motion. This guarantees that the evolution of $A$ is isospectral. In addition to furnishing conserved quantities, having a Lax pair with spectral parameter and $r$-matrix to express the fundamental Poisson brackets, allows for a deeper understanding of the model, say, in terms of its spectral curve \cite{babelon-bernard-talon}.

Lax pairs for integrable oscillators with two degrees of freedom have appeared in the literature before. The most well-known examples are the $N=2$ Toda chain \cite{flaschka}, rational, trigonometric, hyperbolic and elliptic Calogero-Moser-Sutherland models \cite{calogero,sutherland-1971,sutherland-1972,moser}, their relativistic counterparts (Ruijsenaars–Schneider models) \cite{arutyunov-book,hoppe-book} and the Neumann model \cite{avan-talon-neumann-model}. In \cite{kudryashov}, Lax pairs for two coupled nonlinear oscillators were discussed. However, unlike the oscillators we consider, the cited models are generally not circularly symmetric.

We begin in \S \ref{s:lax-pair-iso-harm-osc} with the simplest possibility, the isotropic harmonic oscillator (IHO) on the $x$-$y$ plane. Despite its simplicity, it has interesting features such as all trajectories being periodic due to a third independent conserved quantity, making it one of the simplest (maximally) superintegrable systems.  We recall that it is bi-Hamiltonian: the equations of motion are Hamiltonian with respect to two compatible symplectic forms. Although the recursion operator built from the quotient of symplectic forms gives two of the three conserved quantities and guarantees that they are in involution due to a vanishing Nijenhuis torsion tensor, the corresponding Lax equations are identically satisfied and consequently are not equivalent to the equations of motion. We may try to find a Lax pair for this system by taking the zero coupling limit of a two-particle harmonic Calogero chain. However, we show in \S \ref{s:harmonic-calogero-limit} that a generalized Lax pair for the latter does not admit a nontrivial limit: the Lax matrices commute. Next, in \S \ref{s:4x4-block-lax-pair-iso-harm-osc} we use individual oscillator Lax pairs to construct a conventional $4 \times 4$ block-form Lax pair $(A,B)$ with spectral parameter $\z$. Interestingly, the associated dynamical $r$-matrix $r_{12}(\z, \z')$ depends on only one of the two spectral parameters, the dependence being linear, by contrast to the rational/trigonometric/elliptic $r$-matrices, often depending on $\z - \z'$, encountered elsewhere \cite{babelon-bernard-talon}. However, the spectral invariants of the $4 \times 4$ Lax matrix $A$ still give only two of the three constants of motion. In \S \ref{s:2x2-lax-pair-iso-harm-osc} we find a large class of $2 \times 2$ traceless symmetric-antisymmetric $(A,B)$ pairs with spectral parameter that give all three independent conserved quantities. Our procedure leads to all Lax pairs with $A$ linear in positions and momenta and constant $B$. In a sense, the direct sum of the earlier $4 \times 4$ construction is replaced by an ordinary $\zeta$-weighted sum. Interestingly, not all these Lax pairs are gauge-equivalent under the orthogonal group. However, since the resulting conserved quantities are not in involution (suitably expressed, they obey an $su(2)$ dynamical symmetry algebra), the fundamental Poisson brackets among the Lax matrix entries do not admit an $r$-matrix formulation. Although this construction may be viewed as a warm-up exercise, it perhaps furnishes the simplest example of a Lax pair that leads to all conserved quantities of a superintegrable system.

In \S \ref{s:2x2-antiherm-tr-less-lax-pair-iso-anharm-osc} we move to the circularly symmetric planar oscillator with quartic potential $\al r^2 + \beta r^4$ where $r^2 = x^2 + y^2$. Although we are unable to find a traceless symmetric-antisymmetric $2 \times 2$ Lax pair of the sort that worked in \S \ref{s:2x2-lax-pair-iso-harm-osc}, examination of the equations of motion written in complex position and momentum variables allows us to propose forms for traceless antihermitian $2 \times 2$ Lax matrices $(A,B)$ whose entries are Laurent polynomials in a spectral parameter $\zeta$. By fixing coefficients, we construct a 4-parameter family of Lax pairs whose spectral invariants give both conserved quantities. The fundamental Poisson brackets are encoded in a nondynamical rational $r$-matrix. 

In \S \ref{s:lax-pair-iso-FD-osc-quartic-potn} we generalize these results to an anharmonic oscillator (AHO) with an additional rotational energy governed by the Hamiltonian
    \beq
    H = \frac{p_x^2 + p_y^2}{2\mu} + \frac{\g (x p_y - y p_x)}{\mu} + \al r^2 + \beta r^4.
    \label{e:hamiltonian-FD-osc-quartic-potn}
    \eeq
This is the Fock-Darwin oscillator \cite{fock-1928,darwin-1930} subject to the quartic potential $\beta r^4$. The isotropic Fock-Darwin oscillator ($\beta = 0$) describes a planar charged particle subject to a linear radial electric field and a uniform axial magnetic field. Inspired in part by minimal coupling, we generalize our ansatz from \S \ref{s:2x2-antiherm-tr-less-lax-pair-iso-anharm-osc} to arrive at a 3-parameter family of Lax pairs for this oscillator and a corresponding rational $r$-matrix in \S \ref{s:FD-AHO-FPB-r-matrix-cons-qty}. When $\g \to 0$, these Lax pairs form a subfamily of the ones found in \S \ref{s:2x2-antiherm-tr-less-lax-pair-iso-anharm-osc}. Interestingly, the $\beta \to 0$ limit of the Lax pairs for the nonlinear oscillators of \S \ref{s:2x2-antiherm-tr-less-lax-pair-iso-anharm-osc} and \S \ref{s:lax-pair-iso-FD-osc-quartic-potn} are singular, so they do not reduce to ones for the corresponding isotropic linear oscillators, just as our Lax pairs for the isotropic harmonic oscillator of \S \ref{s:lax-pair-iso-harm-osc} do not reduce to those for the free particle in the limit $\al \to 0$.

In \S \ref{s:compare-RR-model} we show that a change of variables (\ref{e:transformation-LS-intermsof-xypxpy}) relates the Fock-Darwin anharmonic oscillator of (\ref{e:hamiltonian-FD-osc-quartic-potn}) to the equations of the Rajeev-Ranken (RR) model restricted to a symplectic leaf \cite{gsk-trv-rr-mod-anharm-osc}. The RR model \cite{rajeev-ranken,gsk-trv-rr-mod-integrability} is a mechanical system with a six-dimensional phase space that governs the dynamics of screw-type waves in a 1+1 dimensional scalar field theory dual to the SU(2) principal chiral model. Interestingly, this allows us to find a new class of Lax pairs for the RR model going beyond the one given in \cite{gsk-trv-rr-mod-integrability}. What is more, we find associated $r$-matrices for the RR model by transforming those of the Fock-Darwin anharmonic oscillator, although the Poisson structures of the two models are distinct. We conclude in \S \ref{s:discussion} with a brief discussion.

%-----------------
\section{Lax pairs for isotropic harmonic oscillator}
\label{s:lax-pair-iso-harm-osc}
%-----------------

Consider a 2d isotropic linear oscillator with a common mode frequency $\om = \sqrt{2 \al/\mu}$ and equations of motion (EOM)
	\beq
	\mu \dot x = p_x, \quad
	\mu \dot y = p_y, \quad
	\dot p_x = - 2 \al x \quad \text{and} \quad
	\dot p_y = - 2 \al y
	\label{e:EOM-iso-harm-osc-x-y-px-py}
	\eeq
and conserved energy a sum of mode energies
    \beq
    H = E_x + E_y = \frac{p_x^2 + p_y^2}{2 \mu} + \al(x^2 +y^2).
    \label{e:hamil-iso-2d-sho}
    \eeq
This system is bi-Hamiltonian \cite{das-integ-models,ral-fernandes}: it admits two Hamiltonian formulations. If we let $\xi = (x,y,p_x,p_y)$ denote coordinates on phase space, the EOM (\ref{e:EOM-iso-harm-osc-x-y-px-py}) are expressible as $\dot \xi^a = r_1^{ab} \pdr_b H_1 = r_2^{ab} \pdr_b H_2$. The first Hamiltonian ($H_1$) is the energy with canonical Poisson tensor:
    \beq
    H_1 = H = E_x + E_y \quad \text{and} \quad \{ \xi^a, \xi^b \}_1 = r_1^{ab} = \begin{smmat} 0 & I \\ -I & 0 \end{smmat}.
    \eeq
The second Hamiltonian ($H_2$) and Poisson brackets are less familiar:
    \beq
    H_2 = \eps (e^{E_x/\eps} + e^{E_y/\eps}) \quad \text{and} \quad
    \{ \xi^a, \xi^b \}_2 = r_2^{ab} = \begin{smmat} 0 & D \\ -D & 0 \end{smmat} \quad \text{where} \quad
    D = \begin{smmat} e^{-E_x/\eps} & 0 \\ 0 & e^{-E_y/\eps} \end{smmat}.
    \eeq
Here, $\eps$ is an arbitrary fixed parameter with dimensions of energy. For simplicity, we could work in units where $\eps = 1$. What is more, the two brackets are compatible in the sense that their linear combination, the $\la$-bracket $\{ \cdot, \cdot \}_\la = \{ \cdot, \cdot \}_ 1 + \la \{ \cdot , \cdot \}_2$ is also a Poisson bracket for any real $\la$. We have checked that it satisfies the Jacobi identity. Given that the Poisson structures are nondegenerate, we attempt to define a Lax pair $(A,B)$ using Magri's recursion operator \cite{magri} $A = r_2^{-1} r_1$ and $B_a^b = - \pdr_a \dot \xi^b$. We find that $A = \diag(e^{E_x/\eps}, e^{E_y/\eps}, e^{E_x/\eps}, e^{E_y/\eps})$ while $B = \begin{smmat} 0 & 2 \al I \\ -I/\mu & 0 \end{smmat}$. The eigenvalues of $A$ give two independent conserved quantities $E_x$ and $E_y$ and moreover, we have checked that the Nijenhuis tensor $-N^a_{bc} = A^d_b \pdr_d A^a_c - A^d_c \pdr_d A^a_b - A^a_d (\pdr_b A^d_c - \pdr_c A^d_b)$ associated to the $(1,1)$ tensor $A$ vanishes, guaranteeing that the eigenvalues of $A$ are in involution \cite{magri-morosi}. However, the putative Lax equation $\dot A = [B,A]$ is trivially satisfied since both sides vanish identically. Thus, we do not obtain a Lax pair from this bi-Hamiltonian formulation.

%-----------------
\subsection{Limit of the harmonic Calogero model}
\label{s:harmonic-calogero-limit}
%-----------------

Given that Calogero models are known to admit Lax pairs, it is interesting to examine whether one can obtain a Lax pair for the isotropic harmonic oscillator (IHO) by viewing it as a limiting case of a Calogero model. Of relevance to us is the 2-particle harmonic Calogero model (hCM) with Hamiltonian
    \beq
    H = \ov{2\mu}(p_x^2 + p_y^2) + \half \mu \om^2 (x^2 + y^2) + \half \frac{g^2}{(x-y)^2}.
    \eeq
In the limit $g \to 0$, it reduces to the 2d isotropic harmonic oscillator. A generalized Lax pair $(L^\pm, M)$ for the hCM satisfying the Lax equations $\dot L^\pm = -i[M, L^\pm] \pm i \om L^\pm$ was reported in \cite{abanov-gromov-kulkarni}. In the limit $g \to 0$, the Lax matrices become somewhat trivial and the commutator term vanishes:
    \beq
    L^\pm = \diag(p_x \pm i \mu \om x, p_y \pm i \mu \om y) 
    \quad \text{and} \quad M = 0.
% \begin{smmat} p_x \pm i \mu \om x & 0 \cr 0 &  p_y \pm i \mu \om y \end{smmat}
    \eeq
Nevertheless, the surviving equations reproduce Hamilton's equations for the isotropic harmonic oscillator. The resulting conserved quantities are given by $I_k = \tr A^k$ where $A = L^- L^+ = 2 \mu \diag(E_x, E_y)$ with $E_x$ and $E_y$ the mode energies. However, only two of these are independent. In fact, $I_1 = 2 \mu (E_x + E_y)$ is proportional to the total energy while $I_2 = 4 \mu^2 (E_x^2 + E_y^2)$ is an independent integral of the motion. On the other hand, the Cayley-Hamilton formula $A^2 = A \tr A - (\det A) I$ implies that $I_3, I_4, \ldots$ are functions of $I_1$ and $I_2$.

It would be desirable to find a conventional Lax pair for the isotropic harmonic oscillator that depends on a spectral parameter and admits an $r$-matrix formulation which would guarantee that conserved quantities are in involution. We will present such a Lax pair in the next section and follow that up with another one that gives all three conserved quantities.

%-----------------
\subsection{Block-form \texorpdfstring{$4 \times 4$}{4 x 4} Lax pair for isotropic harmonic oscillator}
\label{s:4x4-block-lax-pair-iso-harm-osc}
%-----------------

We construct a Lax pair and $r$-matrix for the 2d isotropic linear oscillator (\ref{e:hamil-iso-2d-sho}) by combining those for 1d oscillators in a suitable block form.

\paragraph{\bf Lax pair and \boldsymbol{$r$}-matrix for 1d harmonic oscillator.} Recall \cite{arutyunov-book,babelon-bernard-talon} that the EOM of a 1d harmonic oscillator $\mu \dot x = p$, $\dot p = - \mu \om^2 x$ are equivalent to the Lax equations $\dot A = [B,A]$ where
	\beq
	A = \colvec{2}{p/\mu & \om x}{\om x & -p/\mu} \quad
	\text{and} \quad
	B = \frac{\om}{2} \colvec{2}{0 & -1}{1 & 0}
	\eeq
are traceless symmetric and antisymmetric matrices acting on the vector space $U = \mathbb{R}^2$. As a consequence, $\tr A^n$ are conserved and in this case $\tr A^2$ is a multiple of the energy. Moreover, the fundamental Poisson brackets (FPBs) between entries of the Lax matrix $\{ A_{12}, A_{11} \} = \om/\mu$ and $\{ A_{12}, A_{22} \} = - \om/\mu$ may be expressed in terms of a classical $r$-matrix  (see Appendix  \ref{a:fpb-r-mat-for-4x4-isotropic-sho-lax} for the notation): 
    \beq
    \{ A_1 \stackrel{\otimes}{,} A_2 \} = [r_{12}, A_1] - [r_{21}, A_2] \quad \text{where} \quad 
    A_1 = A \otimes I \quad \text{and} \quad A_2 = I \otimes A.
    \eeq
To ensure antisymmetry of the FPBs, the linear operators $r_{12}$ and $r_{21}$ that act on $U \otimes U$  must be related by an exchange of spaces: $r_{21} = P r_{12} P^{-1}$ where $P$ is the permutation $P(v \otimes w) = w \otimes v$ for any $v, w \in U$. By the Babelon-Viallet Theorem \cite{babelon-viallet-1990}, the existence of such an $r$-matrix guarantees that the eigenvalues of $A$ are in involution. Alternatively, we may show that the conserved spectral invariants $\tr A^p$ for $p=1,2,3, \ldots$ Poisson commute:
    \beqs
    \{ \tr A^p, \tr A^q \} &=& \dd{\tr A^p}{A_{\al \beta}} \dd{\tr A^q}{A_{\g \del}} \{ A_{\al \beta}, A_{\g \del} \}
    = p A^{p-1}_{\beta \al} q A^{q-1}_{\del \g} \{ A_{\al \beta}, A_{\g \del} \} \cr
    &=& p q \tr ( A_1^{p-1} A_2^{q-1} ([r_{12}, A_1] - [r_{21}, A_2])) = 0.
    \label{e:pb-trAp-trAq-rmat-sans-spec-param}
    \eeqs
The $r_{12}$ and $r_{21}$ terms cancel separately by the cyclicity of the trace and upon using $[A_1, A_2]=0$. For the 1d oscillator with energy $E = (1/2)(p^2/\mu + \mu \om^2 x^2)$, we may take
	\beq
	r_{12} = \ka S \otimes A \quad \text{and} \quad
	r_{21} = \ka A \otimes S \quad \text{where} \quad
	S = \begin{smmat} 0 & 1 \\ -1 & 0 \end{smmat}
	\quad \text{and} \quad
	\ka = - \om /4E.
	\label{e:r-matrix-1d-sho}
	\eeq

\paragraph{\boldsymbol{$4 \times 4$} Lax pair for isotropic harmonic oscillator.} Returning to the 2d oscillator, let us denote $x$ and $y$ by $x_+$ and $x_-$ so that
	\beq
	\mu \dot x_\pm = p_\pm \quad \text{and} \quad
	\dot p_\pm = - \mu \om^2 x_\pm.
	\label{e:EOM-iso-harm-osc-xpm-ppm}
	\eeq
We introduce the $2 \times 2$ traceless symmetric and antisymmetric matrices
	\beq
	A_\pm = \colvec{2}{p_\pm/\mu & \om x_\pm}{ \om x_\pm & -p_\pm/\mu} \quad \text{and} \quad
	B_+ = B_- = \frac{\om}{2} \colvec{2}{ 0 & 1}{-1 & 0 }.
    \label{s:4x4-A_+--B_+--indv-mode}
	\eeq
Then Hamilton's equations (\ref{e:EOM-iso-harm-osc-xpm-ppm}) are equivalent to the Lax equations $\dot A = [B,A]$ for the pair
	\beq
	A = \colvec{2}{ A_+ & 0}{0 & \z A_-} \quad \text{and} \quad 
	B = \colvec{2} { B_+ & 0}{ 0 & B_-}.
	\label{e:4x4-lax-pair-iso-harm-osc}
	\eeq
The introduction of the spectral parameter $\z$ allows us to obtain both conserved mode energies $H_\pm = p_\pm^2/2 \mu + \mu \om^2 x_\pm^2/2$ from this Lax pair. In fact, $\tr A^2 = \tr A_+^2+ \z^2 \tr A_-^2$ is conserved for any $\z$, whence $\tr A_\pm^2 = (4/\mu) H_\pm$ are individually conserved. Due to the block structure, no additional independent conserved quantity can arise from traces of higher powers of $A$. In particular, the third conserved quantity ($L_z = x p_y - y p_x$) does not follow from this Lax pair. We will return to this issue in \S \ref{s:2x2-lax-pair-iso-harm-osc} after constructing an $r$-matrix for this Lax pair.

%---------------
\subsubsection{Fundamental Poisson brackets and \texorpdfstring{$r$}{r}-matrix}
\label{s:4x4-fpb-r-matrix}
%---------------

The equations of motion (\ref{e:EOM-iso-harm-osc-xpm-ppm}) are Hamiltonian with respect to the canonical Poisson brackets (PBs) $\{ x_i, p_j \} = \del_{ij}, \{ x_i, x_j \} = \{ p_i, p_j \} = 0$ where $i,j \in \{+,- \}$. Consequently, the conserved energies $H_\pm$ commute. Noting that $A$ is symmetric, the independent nonzero FPBs between entries of $A$ (up to antisymmetry) are
	\beqs
	\{A_{12}(\z), A_{11}(\z')\} =
	- \{A_{12}(\z), A_{22}(\z')\} &=& \om/\mu,
	\cr \text{and} \qquad
	\{A_{34}(\z), A_{33}(\z')\} =
	- \{A_{34}(\z), A_{44}(\z')\} &=& \z \z' \om/\mu.
	\label{e:fpb-spec-para-2d-sho}
	\eeqs
To find an $r$-matrix formulation 
	\beq
	\{ A_1(\z), A_2(\z') \} = [r_{12}(\z, \z'), A_1(\z)] - [r_{21}(\z, \z'), A_2(\z')]
	\label{e:fpb-rmatrix-spec-para}
	\eeq
of these FPBs, we will use a direct sum decomposition of the auxiliary vector spaces $V$ and $W = V \otimes V$ on which $A$ and the $r$-matrix act. The details of this construction are given in Appendix \ref{a:fpb-r-mat-for-4x4-isotropic-sho-lax} and lead us to the block diagonal $r$-matrix:
    \beqs
    r_{12}(\z,\z') &=& \diag(\ka_+ S \otimes A_+, \: \z' \ka_- S \otimes A_-, \: 0, \: 0) \quad \text{and} \cr
    r_{21}(\z,\z') &=& \diag(\ka_+ A_+ \otimes S, \; \z \ka_- A_- \otimes S, \: 0, \: 0).
    \eeqs
Here, $A_\pm$ (\ref{s:4x4-A_+--B_+--indv-mode}) are the individual mode Lax matrices and $S = \begin{smmat} 0 & 1 \cr -1 & 0 \end{smmat}$. This $r$-matrix is dynamical: the mode energies enter through $\ka_\pm = -\om/4 H_\pm$. Moreover, $r_{12}(\z,\z')$ depends on only one of the spectral parameters ($\z$) and not, as is more common, on the difference $\z -\z'$. Its existence ensures that the conserved energies $H_\pm$ commute. In fact, from (\ref{e:fpb-rmatrix-spec-para}), proceeding as in (\ref{e:pb-trAp-trAq-rmat-sans-spec-param}), we show that \small
    \beqs
    \{\tr A^p(\z), \tr A^q(\z^\pr) \} 
    &=& 
    \iint d \z'' d \z''' \dd{\tr A^p(\z)}{A_{\al \beta}(\z'')} \dd{\tr A^q(\z')}{A_{\g \del}(\z''')} \{A(\z'')_{\al \beta} , A(\z''')_{\g \del} \}  \cr
    &=& \dd{\tr A^p(\z)}{A_{\al \beta}(\z)} \dd{\tr A^q(\z')}{A_{\g \del}(\z')} \{A(\z)_{\al \beta} , A(\z')_{\g \del} \} 
    \cr
    &=& pq \tr A_1(\z)^{p-1} A_2(\z')^{q-1} ([r_{12}(\z, \z^\pr), A_1(\z)] - [r_{21}(\z, \z'), A_2(\z')]) = 0.
    \quad
    \label{e:trAp-trAq-pb-with-spectralparam}
    \eeqs \normalsize
In the last step, we use the cyclicity of $\tr$ and the fact that $A_1(\z)$ commutes with $A_2(\z')$.

%-----------------
 \subsection{Symmetric-anti-symmetric \texorpdfstring{$2 \times 2$}{2 x 2} Lax pairs for isotropic harmonic oscillator}
\label{s:2x2-lax-pair-iso-harm-osc}
%-----------------

\fl {\bf Ansatz for Lax matrices.} The $4 \times 4$ block diagonal Lax pair (\ref{e:4x4-lax-pair-iso-harm-osc}) of \S \ref{s:4x4-block-lax-pair-iso-harm-osc} leads to the two conserved mode energies of the isotropic 2d harmonic oscillator. However, being superintegrable, the system has a third independent conserved quantity, which can be taken as the $z$-component of angular momentum $L_z = x p_y - y p_x$. We now look for a Lax pair that gives all three conserved quantities. In fact, we will realize Hamilton's equations
	\beq
	\mu \dot x = p_x, \quad
	\mu \dot y = p_y, \quad
	\dot p_x = - 2 \al x \quad \text{and} \quad
	\dot p_y = - 2 \al y,
	\label{e:EOM-iso-harm-osc-x-y-px-py-2nd-time}
	\eeq
for an isotropic oscillator with frequency $\om = \sqrt{2 \al/\mu}$ as the Lax equations $\dot A = [B,A]$ for $2 \times 2$ matrices. We take $A = (U, V | V, -U)$ to be a real symmetric traceless matrix so that it has two independent entries. As we have four first order EOM, we suppose that $U$ and $V$ involve terms of order $\z^0$ and $\z^1$ in a spectral parameter $\z$. Since the EOM are linear, we take $U$ and $V$ to be linear in $x, y, p_x, p_y$ and $B$ to be independent of dynamical variables. For consistency of the Lax equation $\dot A = [B,A]$, the commutator must be symmetric. To achieve this, we chose $B$ to be antisymmetric. Any symmetric part of $B$ will not contribute to $[B,A]$ if the latter is symmetric. Thus, we begin with the ansatz
	\beq
	A = \colvec{2}{U & V}{V & -U},
%	= V \sig_1 + U \sig_3,
	\;\; B = \nu \colvec{2}{0 & -1}{1 & 0} \;\; \text{where} \;\;
	U = (a + \z a') \cdot \xi \quad 
	\text{and} \quad
	V = (b + \z b') \cdot \xi.
	\label{e:lax-pair-2x2-symm-antisymm-str}
	\eeq
The positions, momenta and real coefficients have been arranged in vectors:
	\beq
	\xi = (x, y, p_x, p_y), \quad
	a = (a_1, a_2, a_3, a_4), \quad
	b = (b_1, b_2, b_3, b_4), \quad \text{etc.}
	\eeq
The Lax equation
	\beq
    	\dot A = \colvec{2}{\dot U & \dot V}{\dot V & -\dot U} = [B,A] = 2 \nu \colvec{2}{- V & U}{U & V}
	\eeq
becomes a pair of first order equations for $U$ and $V$
	\beq
	\dot U = -2 \nu V \quad \text{and} \quad
	\dot V = 2 \nu U.
	\label{e:Lax-eqn-U-V} 
	\eeq 
Isolating the coefficients of $\z^0$ and $\z^1$ we get four Lax equations that we order as
	\beq
	a \cdot \dot \xi = -2 \nu b \cdot \xi, \quad
	a' \cdot \dot \xi = -2 \nu b' \cdot \xi, \quad
	b \cdot \dot \xi = 2 \nu a \cdot \xi, \quad
	b' \cdot \dot \xi = 2 \nu a' \cdot \xi.
	\label{e:4-lax-eq-in-order}
	\eeq
Similarly, we have chosen an order for Hamilton's equations in (\ref{e:EOM-iso-harm-osc-x-y-px-py-2nd-time}). We ask whether $\nu$ and the 16 coefficients $a, a', b, b'$ may be chosen so that the four Lax equations are equivalent to the EOM in this order. It turns out this is possible. More generally, there are $4!$ ways of ordering Hamilton's equations. Of these, we find that 16 orderings are inconsistent with the Lax equations while the remaining eight lead to 8 families of Lax pairs.

\vspace{4pt}

\fl {\bf Lax pair for $\boldsymbol{x,y,p_x,p_y}$ ordering of EOM.} To see this, let us begin by comparing the first EOM $\mu \dot x = p_x$ with the first Lax equation $a \cdot \dot \xi = - 2 \nu b \cdot \xi$.
%	\beq
%	a_1 \dot x + a_2 \dot y + a_3 \dot p_x + a_4 \dot p_y = - 2 \nu (b_1 x + b_2 y + b_3 p_x + b_4 p_y).
%	\eeq
They are equivalent iff $a_1 = - 2 \mu \nu b_3$ with the remaining coefficients vanishing. Comparing the other EOM with the corresponding Lax equations, we get the conditions
	\beq
	a_2' = - 2 \mu \nu b_4', \quad
	\al b_3 = - \nu a_1 \quad
	\text{and} \quad
	\al b_4' = - \nu a_2',
	\eeq
while the remaining coefficients must vanish. Combining, we find the nonvanishing coefficients
	\beq
	\nu = \pm \om/2, \quad
	a_1 = \mp \mu \om b_3 \quad \text{and} \quad
	a_2' = \mp \mu \om b_4'.
	\eeq
It follows that
	\beq
	U = \mp \mu \om (b_3 x + \z b_4' y) 
	\quad \text{and} \quad
	V = b_3 p_x + \z b_4' p_y,
	\eeq
where $b_3, b_4'$ are arbitrary nonzero constants, for simplicity we will take them equal to unity.
This leads to the Lax pair:
	\beq
	A_1 = \colvec{2}{\mp \mu \om (x + \z y) &  p_x + \z p_y}{p_x + \z p_y & \pm \mu \om (x + \z y)} \quad
	\text{and} \quad
	B = \mp \frac{\om}{2} \colvec{2}{0 & 1}{-1 & 0}.
	\label{e:lax-1}
	\eeq
We notice that $A_1$ is a $\zeta$-weighted linear combination of Lax matrices for the $x$ and $y$ variables: $A_1 = b_3 A_{1x} + \z b'_4 A_{1y}$ where $(A_{1x}, B)$ and $(A_{1y}, B)$ are Lax pairs for the individual oscillators. This additive feature of Lax matrices for a common $B$ may be traced to the isotropy of the oscillator.

\vspace{4pt}

\fl {\bf Orderings that do not lead to a Lax pair.} Let us next consider an ordering of the EOM 
	\beq
	\mu \dot x = p_x, \quad
	\dot p_x = - 2 \al x, \quad
	\mu \dot y = p_y, \quad
	\dot p_y = - 2 \al y,
	\eeq
that turns out to be inconsistent when matched with the Lax equations as ordered in (\ref{e:4-lax-eq-in-order}).   Upon comparing, we get the conditions
	\beqs
	a_1 = - 2 \mu \nu b_3, && a_2 = a_3 = a_4 = b_1 = b_2 = b_4 = 0, \cr
	\al a_3' = \nu b_1', && a_1' = a_2' = a_4' = b_2' = b_3' = b_4' = 0, \cr
	b_2 = 2 \mu \nu a_4, && a_1 = a_2 = a_3 = b_1 = b_3 = b_4 = 0, \cr
	- \al b_4' = \nu a_2', && a_1' = a_3' = a_4' = b_1' = b_2' = b_3' = 0.
	\eeqs
Nontrivial solutions are not possible due to incompatibilities between the $1^{\rm st}$ and $3^{\rm rd}$ as well as between the $2^{\rm nd}$ and $4^{\rm th}$ equations.

As one might suspect and infer from the above examples, to avoid such inconsistencies, it is necessary and sufficient that both the $x$ oscillator equations be matched with either the $\z^0$ or $\z^1$ Lax equations (i.e., both must involve either unprimed or primed coefficients) while the $y$ oscillator equations be matched with the other pair of Lax equations. There are 8 such consistent matchings leading to 8 types of Lax pairs, the first of which was given in (\ref{e:lax-1}). 

\vspace{4pt}

\fl {\bf Swapped and mixed Lax pairs from other orderings.} A second possibility comes from the order
	\beq
	\dot p_x = - 2 \al x, \quad
	\dot p_y = - 2 \al y, \quad
	\mu \dot x = p_x, \quad
	\mu \dot y = p_y,
	\eeq
which essentially leads to a reversal of roles played by positions and momenta in the Lax pair. In fact, we get the following conditions $\al a_3 = \nu b_1$, $\al a_4' = \nu b_2'$, $b_1 = 2 \mu \nu a_3$ and $b_2' = 2 \mu \nu a_4'$ while the remaining coefficients must vanish. The nontrivial coefficients are 
    \beq
    \nu = \pm \om/2, \quad b_1 = \pm \mu \om a_3 \quad \text{and} \quad b_2' = \pm \mu \om a_4',
    \eeq
leading to $U = a_3 p_x + \z a'_4 p_y$ and $V = \pm \mu \om (a_3 x + \z a_4' y)$. Taking $a_3 = a_4' = 1$, we get
	\beq
	A_2 = \colvec{2}{p_x + \z p_y & \pm\mu \om(x + \z y)}{\pm \mu \om (x + \z y) & -p_x - \z p_y} \quad
	\text{and} \quad
	B = \mp \frac{\om}{2} \colvec{2}{0 & 1}{-1 & 0}.
	\label{e:lax-2}
	\eeq
On the other hand, the order $\mu \dot x = p_x,\dot p_y = - 2 \al y$, $\dot p_x = -2 \al x, \mu \dot y = p_y$ leads to a `mixed' Lax pair:
	\beq
	A_3 = \colvec{2}{\mp \mu \om x + \z p_y & p_x \pm \z \mu \om y}{p_x \pm \z \mu \om y & \pm \mu \om x - \z p_y}
	\quad
	\text{and} \quad
	B = \mp \frac{\om}{2} \colvec{2}{0 & 1}{-1 & 0}.
	\label{e:lax-3}
	\eeq
As with $A_1$, $A_2$ and $A_3$ are linear combinations of Lax matrices of individual oscillators weighted by $\z$. In a sense, the direct sum in (\ref{e:4x4-lax-pair-iso-harm-osc}) is replaced here by an ordinary sum.

\vspace{4pt}

\fl {\bf Gauge transformation between Lax matrices.} It is noteworthy that $A_2$ is related to $A_1$ via a `gauge transformation', i.e., conjugation by an orthogonal matrix (to preserve tracelessness): $A_2 = g A_1 g^t$ where $g = \ov{\sqrt{2}} \begin{smmat} 1 & 1 \cr -1 & 1 \end{smmat}$. However, $A_3 \ne g A_1 g^t$ for any orthogonal $g$ whose entries could be functions on the phase space. In other words, the group of gauge transformations does not act transitively on the space of such $2 \times 2$ Lax matrices.

%-----------------
\subsubsection{All three conserved quantities from Lax pairs and Fundamental Poisson brackets}
%-----------------

A novel feature of the Lax pairs for the 2d isotropic harmonic oscillator obtained in \S \ref{s:2x2-lax-pair-iso-harm-osc} is that they give all three independent conserved quantities, not just two that are in involution. Observe that the Lax matrix $A$ (\ref{e:lax-pair-2x2-symm-antisymm-str}) is traceless, while $\tr A^2 = 2(U^2 + V^2)$ is a quadratic polynomial in $\z$, so each coefficient is conserved. For instance,
	\beq
	\tr A_1^2 = 4 \mu (H_{xx} + \z H_{xy} + \z^2 H_{yy}) 
	\quad \text{and} \quad
	\tr A_3^2 = 4 \mu (H_{xx} \mp \z \om L_z + \z^2 H_{yy}).
	\label{e:tr-lax1-lax3-sq-3-cons-qty}
	\eeq
Here, $L_z = x p_y - y p_x$ while $H_{xx} = E_x$ and $H_{yy} = E_y$ are the mode energies from (\ref{e:hamil-iso-2d-sho}). $H_{xy} = p_x p_y/\mu + \mu \om^2 xy$ is a third independent component of the conserved Fradkin tensor $F_{ij} = p_i p_j/\mu + \mu \om^2 x_i x_j$. They satisfy the quadratic relation
	\beq
	H_{xy}^2 = 4 H_{xx} H_{yy} - \om^2 L_z^2.
	\eeq
What is more, the wedge product $dL_z \wedge d H_{xx} \wedge dH_{yy}$ is generally nonzero on the phase space. Thus, any three of these four are functionally independent conserved quantities. 

% $A^2 = (U^2 + V^2)I$ where $U$ and $V$ are linear in $\z$.

\paragraph{Fundamental Poisson brackets and algebra of conserved quantities.} Taking $(x, p_x)$ and $(y, p_y)$ to be canonically conjugate pairs we may obtain the fundamental PBs among entries of the Lax matrix. If we let $\phi(\z, \z') = \mp \mu \om (1 + \z' \z)$, then from (\ref{e:lax-1}) the nonvanishing FPBs (up to antisymmetry) for $A_1$ are:
	\beqs
	\{A_{11}(\z), A_{12}(\z')\} = \{A_{11}(\z), A_{21}(\z')\} &=& \phi(\z, \z' ),
	\cr
	\{A_{22}(\z), A_{12}(\z')\} =  \{A_{22}(\z), A_{21}(\z')\} &=& -\phi(\z, \z').
	\label{e:fbp-2x2-lax-2d-isotropic-sho}
	\eeqs
The Poisson algebra of any three of the above conserved quantities (say, $H_{xx}, H_{yy}$ and $H_{xy}$) is nonlinear but may be made to look linear by including a fourth (say, $L_z$):
	\beqs
	\{ H_{xx}, H_{yy} \} &=& 0, \quad
	\{ H_{xx}, H_{xy} \} = - \{ H_{yy}, H_{xy} \}  = (\om^2/\mu) L_z, \cr
	\{ H_{xx}, L_z \} &=& - \{ H_{yy}, L_z \} = - H_{xy}
	\quad \text{and} \quad
	\{ H_{xy}, L_z \} = 2 (H_{xx} - H_{yy}).
	\eeqs
In particular, the conserved quantities arising as coefficients of distinct powers of $\z$ in $\tr A^2$ (\ref{e:tr-lax1-lax3-sq-3-cons-qty}) are not in involution: at most two can commute for two degrees of freedom. As is well-known, this nonlinear Poisson algebra can be expressed as the $su(2)$ Lie algebra $\{ J_a, J_b \} = \eps_{abc} J_c$ if we introduce
    \beq
    J_1 = \sqrt{I_x I_y} \cos(\D \tht), \quad
    J_2 = \sqrt{I_x I_y} \sin(\D \tht) \quad \text{and} \quad
    J_3 = \half (I_x - I_y).
    \eeq
Here $(I_x, \tht_x)$ and $(I_y, \tht_y)$ are action-angle variables for the $x$ and $y$ modes and $\D \tht = \tht_x - \tht_y$ is a third conserved quantity:
    \beq
    I_x = H_{xx}/\om, \quad
    \tht_x = \arctan(\mu \om x/p_x) \quad \text{and} \quad
    x \leftrightarrow y.
    \eeq
The nonabelian nature of this algebra of conserved quantities implies that the FPBs (\ref{e:fbp-2x2-lax-2d-isotropic-sho}) cannot be expressed as a commutator with a conventional $r$-matrix. We wonder whether there is some way of writing the FPBs that encodes this nonabelian Poisson algebra.

%-----------------
\section{Traceless antihermitian \texorpdfstring{$2 \times 2$}{2 x 2} Lax pair for isotropic anharmonic oscillator}
\label{s:2x2-antiherm-tr-less-lax-pair-iso-anharm-osc}
%-----------------

We next consider an isotropic quartic anharmonic oscillator (AHO) governed by canonical Poisson brackets $\{ x, p_x \} = \{y, p_y \} = 1$ and the Hamiltonian 
    \beq
    H = (1/2\mu) (p_x^2 + p_y^2) + \al r^2
    + \beta r^4 \quad \text{where} \quad r^2 = x^2 + y^2.
    \label{e:hamil-quad+quart-anharm-osc}
    \eeq
The force constant $\al$ and quartic coupling $\beta$ have dimensions $M/T^2$ and $M/L^2 T^2$. Hamilton's equations are
	\beq
 	\mu \dot x = p_x, \quad 
	\mu \dot y = p_y, \quad 
	\dot p_x = - 2 (\al + 2 \beta r^2) x
	\quad  \text{and} \quad
	\dot p_y = - 2 (\al + 2 \beta r^2) y.
	\label{e:rr-mod-eom-beta-nonzero}
	\eeq
% While $\mu$ is a mass, the harmonic and anharmonic couplings have the dimensions $[\al] = M/T^2$ and $\beta = M/L^2 T^2$. 

\vspace{4pt}

\fl {\bf Ansatz for the AHO Lax pair.} We searched for a real symmetric-antisymmetric $2 \times 2$ Lax pair (as in \S \ref{s:2x2-lax-pair-iso-harm-osc}) for this system but did not find one. Here, we will show that (\ref{e:rr-mod-eom-beta-nonzero}) are equivalent to the equations $\dot A = [B, A]$ for a pair of traceless antihermitian matrices whose entries are Laurent polynomials in a spectral parameter $\zeta$. We write
    \beq
    A(\z) = \colvec{2}{U(\z) & V(\z)}{-\bar V(\z) & \bar U(\z)} \quad \text{and} \quad
    B(\z) = \colvec{2}{X(\z) & Y(\z)}{-\bar Y(\z) & \bar X(\z)}.
    \label{e:A-B-antiherm-lax-pair-UVXY}
    \eeq
For real $\zeta$, $U$ and $X$ are imaginary while $V$ and $Y$ are generally complex. In expressions such as $\bar V(\z)$ obtained from (\ref{e:V-ansatz-2d-anharm-osc}), the coefficients are conjugated but $\z$ is not. The Lax equation becomes the pair 
	\beq
	\dot U = V \bar Y - \bar V Y \quad \text{and} 
	\quad \dot V = 2(VX - UY).
	\label{e:lax-eq-U-dot-V-dot-2x2-antiherm}
	\eeq
% Thus, $X$ must have dimensions of $1/T$. Since, $\tr A^2 = 2 (U^2 - V \bar V)$ must be conserved, $U$, $V$ and $A$ must all have a common dimension. This implies that $[Y] = [X] = 1/T$. The freedom to rescale $A$ without altering the Lax equation implies we are free to choose its dimension. We will find a large family of Lax pairs of this sort with various dimensions for $\zeta$.
It is convenient to express the EOM in terms of the complex dynamical variables $Z = x + i y$ and $P = p_x + i p_y$. In fact, (\ref{e:rr-mod-eom-beta-nonzero}) are the real and imaginary parts of
	\beq
	\mu \dot Z = P \quad \text{and} \quad 
	\dot P = - 2(\al + 2 \beta |Z|^2)Z.
	\label{e:Z-dot-P-dot-complex-RR-eom-gamma=0}
	\eeq
We would like to realize (\ref{e:Z-dot-P-dot-complex-RR-eom-gamma=0}) as the Lax equations for $\dot U$ and $\dot V$ in (\ref{e:lax-eq-U-dot-V-dot-2x2-antiherm}) at suitable orders in $\zeta$. Since $V$ is complex, it is natural to take it to be a linear combination of $Z$ and $P$ while $U$ being purely imaginary cannot have this form. Thus, we suppose that 
	\beq
	V = a \zeta Z + b P \quad \text{where $a$ and $b$ are nonzero complex coefficients}
	\label{e:V-ansatz-2d-anharm-osc}
	\eeq
depending on $\mu, \al$ and $\beta$. For definiteness, we have supposed that the $\dot P$ and $\dot Z$ equations arise at $\calO(\zeta^0)$ and $\calO(\zeta^1)$ in the $\dot V$ Lax equation.  The idea is that the $\dot U$ Lax equation will not be an independent equation of motion, but a consequence of the $\dot Z$ and $\dot P$ EOM. Given the form of $V$, we will now try to infer those of $X, Y$ and $U$ by requiring that the Lax equations be equivalent to the EOM.

% So far, we have not fixed the dimension of $\zeta$ (below, $[\zeta] = L$). 

Since we want the Lax equation 
	\beq
	\dot V = a \z \dot Z + b \dot P = 2(VX - UY)
	\label{e:V-dot-lax-eqn-a-b-coeff}
	\eeq
to lead to the $\dot Z$ and $\dot P$ EOM (\ref{e:Z-dot-P-dot-complex-RR-eom-gamma=0}) at order $\z^1$ and $\z^0$, by focusing on the $VX$ term, we are led to suppose that the purely imaginary $X$ is of the form 
	\beq
	X = c \zeta + \frac{d \psi}{\z} \quad \text{where} \quad \psi = \al + 2 \beta |Z|^2.
	\label{e:X-ansatz-2d-anharm-osc}
	\eeq
Here, $c$ and $d$ are imaginary coefficients depending on $\mu, \al$ and $\beta$. It is hoped that the $\calO(1/\zeta)$ and $\calO(\zeta^2)$ terms in $VX$ will cancel similar terms from $UY$ so that they do not contribute to $\dot V$. In a similar vein, the complex $Y$ and the imaginary $U$ must be chosen so that $UY$ contributes at order $\zeta$ and $\zeta^0$ to the $\dot Z$ and $\dot P$ EOM derived from the $\dot V$ Lax equation. This suggests forms similar to those for $V$ and $X$:
	\beq
	Y = \zeta^{-\del} (e \zeta Z + f P)
	\quad \text{and} \quad
	U = \zeta^{\del} \left( g \zeta + j + \frac{h \psi}{\zeta} \right), 
	\label{e:Y-U-ansatz-2d-anharm-osc}
	\eeq
with $e$ and $f$ complex and $g, h$ and $j$ imaginary constants. 

\vspace{4pt}

\fl {\bf Comparing Hamilton and Lax equations to fix coefficients.} We begin by noticing that if $j \ne 0$, then $e=f=c=d=0$ so that $a=b=0$, contradicting (\ref{e:V-ansatz-2d-anharm-osc}). Thus, we take $j = 0$. The pre-factors $\z^{\pm \del}$ cancel out in $UY$, so let us first consider the $\dot V$ Lax equation (\ref{e:V-dot-lax-eqn-a-b-coeff}). Later, we will fix $\del$ by considering the $\dot U$ Lax equation. We find
	\beq
	VX - UY = (a c - e g) \z^2 Z + (b c - f g) \z P + (a d - e h) \psi Z + (b d - f h) \frac{P \psi}{\z} .
	\eeq
Upon comparing with $\dot V$, the coefficients of $\z^2$ and $1/\z$ must vanish: 
	\beq
	ac = eg \quad \text{and} \quad bd = fh.
	\label{e:cond-ac=eg-bd=fh}
	\eeq
The nontrivial Lax equations at $\calO(\z)$ and $\calO(\z^0)$ are:
	\beq
	a \dot Z = 2 (bc - fg) P \quad
	\text{and} \quad
	b \dot P = 2 (ad - eh) \psi Z.
	\eeq
Upon comparing with the EOM (\ref{e:Z-dot-P-dot-complex-RR-eom-gamma=0}) we get two more conditions
	\beq
	2 \mu(bc - fg) = a \quad \text{and} \quad
	eh - ad = b.
	\label{e:3rd-4th-cond-V-dot-lax-eq}
	\eeq

To fix $\del$ and get additional conditions on the coefficients, we consider the first Lax equation  $\dot U = V \bar Y - \bar V Y$. Now 
	\beq
	\dot U = 4 h \beta \z^{\del - 1} r \dot r
	= \frac{4 h \beta}{\mu} \z^{\del - 1} (x p_x + y p_y)
	= \frac{4 h \beta}{\mu} \z^{\del - 1} \Re \bar Z P
	\quad
	\text{since} \quad |Z|^2 = r^2 = x^2 + y^2.
	\label{e:U-dot-RR-model}
	\eeq
On the other hand,
	\beq
	V \bar Y = \z^{-\del} [a \bar e \zeta^2 |Z|^2 + \zeta (a \bar f Z \bar P + b \bar e \bar Z P) + b \bar f |P|^2],
	\eeq
which implies
	\beq
	V \bar Y - \bar V Y = \z^{-\del} [\z^2 |Z|^2 (a \bar e - \bar a e) 
	+ \z ((a \bar f - e \bar b) Z \bar P - (\bar a f - \bar e b) \bar Z P) 
	+ (b \bar f - \bar b f) |P|^2 ].
	\eeq
For $\dot U = V \bar Y - \bar V Y$ to match the EOM $\mu \dot r = p_r$ several conditions must be met: 
\begin{enumerate}
	\item[(i)] the coefficients of $|Z|^2$ and $|P|^2$ must vanish: $\Im a \bar e = 0$ and $\Im b \bar f = 0$; 
	
	\item[(ii)] $\del - 1 = -\del + 1$ or $\del = 1$ and 
	
	\item[(iii)] $(4 h \beta/\mu) \Re \bar Z P = 2 i \Im \{(a \bar f - e \bar b) Z \bar P \}$ for all $Z$ and $P$. 
\end{enumerate}

The conditions $\Im a \bar e = 0$ and $\Im b \bar f = 0$ (i.e., $\bma \times \bme = 0 $, $\bmb \times \bmf = 0$) for nonzero $a, b$ are equivalent to $e = \ka_1 a$ and $f = \ka_2 b$ for some real numbers $\ka_1, \ka_2$. Then $a \bar f - e \bar b = a \bar b (\ka_2 - \ka_1)$. Since $\Im i Z \bar P = \Re \bar Z P$, (iii) implies that $a \bar f - e \bar b$ must be imaginary. Hence, $a \bar b$ must be imaginary and nonzero. This means $b = i \ka_3 a$ for some nonzero real $\ka_3$. This allows us to write $a \bar f - e \bar b = i \ka_3 (\ka_1 - \ka_2) |a|^2$. Condition (iii) then becomes
	\beq
	4 h \beta = 2 i \mu \ka_3 (\ka_1 - \ka_2) |a|^2.
	\eeq
Using the expressions $b = i \ka_3 a$, $e = \ka_1 a$ and $f = i \ka_2 \ka_3 a$, the conditions $ac = eg$ and $bd = fh$ (\ref{e:cond-ac=eg-bd=fh}) become $c = \ka_1 g$ and $d = \ka_2 h$. The remaining two conditions from (\ref{e:3rd-4th-cond-V-dot-lax-eq}) become
	\beq
	2 i \mu \ka_3(\ka_1 - \ka_2) g = 1 \quad
	\text{and} \quad
	(\ka_1 - \ka_2) h = i \ka_3.
	\label{e:last-2-conditions}
	\eeq
Since $\ka_3 \ne 0$, we must have $h \ne 0$ and $\ka_1 \ne \ka_2$. Eliminating $h = i \ka_3/(\ka_1 - \ka_2)$, we are left with two conditions
	\beq
	2 \beta = \mu(\ka_1 - \ka_2)^2 |a|^2 \quad
	\text{and} \quad
	2 i \mu \ka_3(\ka_1 - \ka_2) g = 1.
	\eeq
Eliminating $\ka_1 - \ka_2 = (2i\mu g \ka_3)^{-1}$ we end up with a single condition
	\beq
	|a|^2 + 8 \beta \mu g^2 \ka_3^2 = 0.
	\label{e:a-g-la3-condn}
	\eeq
Thus, we are free to pick any imaginary coefficient $g$, real factor $\ka_3 = b/ia$ and phase for $a$. Eq. (\ref{e:a-g-la3-condn}) then fixes the magnitude of $a$ while (\ref{e:last-2-conditions}) determines the difference $\ka_1 - \ka_2$. {\it Thus, we should expect a 4 parameter ($g, \ka_1, \ka_3, \arg a$) family of Lax pairs of this sort.} Note that $\al$ does not appear in these conditions, the dependence on $\al$ has been incorporated in $X$ and $U$.

\vspace{4pt}

\fl {\bf Example.} We seek a simple Lax pair of the above sort by taking as many coefficients to vanish as is allowed. To begin with, we notice that for $a,b \neq 0$, $g,h$ must also be nonzero because $\ka_3 \neq 0$. However, we can take either $\ka_1$ or $\ka_2$ to vanish. The former would give $e = c = 0$ and the latter would give $d = f = 0$. Since (\ref{e:a-g-la3-condn}) only involves $|a|$, a simple option is to choose $a$ to be real. For instance, (\ref{e:a-g-la3-condn}) can be satisfied by choosing $a = 4 \beta$, $g = 4 i \beta$ and $\ka_3 = 1/\sqrt{8 \beta \mu}$. Moreover, suppose we pick $\ka_1 = 0$, then $\ka_2 = \ka_3$ leading to the following $8$ coefficients:
	\beq
	a = 4 \beta, \;\; 
	b = i \sqrt{2 \beta/\mu}, \;\;
	c = 0, \;\; 
	d = -i/\sqrt{8 \beta \mu}, \;\;
	e = 0, \;\;
	f = i/2\mu, \;\;
	g = 4 i \beta, \;\;
	h = -i.
	\label{e:coeffs-simple-choice-gamma=0}
	\eeq
The matrix elements of the resulting Lax pair (\ref{e:A-B-antiherm-lax-pair-UVXY}) are
	\beqs
	U &=& i (4 \beta \zeta^2 - \psi), \quad
	V = 4 \beta \zeta Z + i \sqrt{2 \beta/\mu} P \quad \text{and} 
	\cr
	X &=& -\frac{i \psi}{\zeta \sqrt{8 \beta \mu}} , \quad
	Y = \frac{i P}{2 \zeta \mu}  \quad 
        \text{where} \quad Z = x + iy, \:\: P = p_x + i p_y.
    \label{e:UVXY-for-AHO-lax-pair}
	\eeqs
For this choice of parameters, $\z$ has dimensions of length. Unfortunately, $\beta \to 0$ is a singular limit that prevents us from obtaining a limiting Lax pair for the linear oscillator. On the other hand, taking $\al \to 0$ (in $\psi$), we get a Lax pair for the quartic oscillator without any linear restoring force.

% $[A] = M/T^2$

 \iffalse
Another possible choice is
	\beq
	a = 1, \;\; 
	b = i, \;\;
	c = -i/2 \mu, \;\; 
	d = 0, \;\;
	e = -\sqrt{2 \beta/\mu}, \;\;
	f = 0, \;\;
	g = i/\sqrt{8 \beta \mu}, \;\;
	h = -i \sqrt{\mu/ 2 \beta},
	\eeq
where we have taken $\ka_1 = -\sqrt{2 \beta/\mu}$, $\ka_2 = 0$ and $\ka_3 = 1$ resulting in $[\z] = M/T$.
 \fi

\vspace{4pt}

\fl {\bf Conserved quantities.} The conserved quantities arise as coefficients of powers of $\z$ in $\tr A^2 = 2 (U^2 - \bar V V)$. Using $\Re (i Z \bar P) = L_z$ we find
%    \beq
%    U^2 = g^2 \z^4 + 2 g h \psi \z^2 + h^2 \psi^2, \quad
%    \bar V V = |a|^2 |Z|^2 \z^2 - 2 \ka_3 |a|^2 \z L_z + |b|^2 |P|^2
%    \eeq
% so that
    \beq
    U^2 - \bar V V = g^2 \z^4 + 2 \al g h \z^2 + 2 \ka_3 |a|^2 L_z \z - \frac{\mu \al^2 \ka_3^2 |a|^2}{2 \beta} - 2 \mu \ka_3^2 |a|^2 \left( |P|^2/2\mu + \al |Z|^2 + \beta |Z|^4 \right).
    \eeq
From (\ref{e:hamil-quad+quart-anharm-osc}), the factor in parentheses is the Hamiltonian $H$. So it follows that
    \beq
    \tr A^2 = 2 g^2 \left( \z^4 - 4 \mu \al \ka_3^2 \z^2 - 16 \mu \beta \ka_3^3 L_z \z + 4 \mu^2 \al^2 \ka_3^4 + 16 \mu^2 \beta \ka_3^4 H \right).
    \eeq
Up to constants, the conserved quantities $L_z$ and $H$ appear as coefficients of $\z^1$ and $\z^0$.

\vspace{8pt}

\fl {\bf Fundamental PBs and \boldsymbol{$r$}-matrices.} Using $\{Z,P\} =  \{Z, \bar Z\} = \{P, \bar P \} = 0$, $ \{\bar Z, P\} = 2$ and $\{ \psi, Z \} = 0$ and $\{ \psi, P \} = 4 \beta Z$ and their complex conjugates, we get  the FPBs between entries of $A(\z) = (U, V | - \bar V, \bar U)$ (\ref{e:UVXY-for-AHO-lax-pair}): \small
    \beqs
    \{U(\z), U(\z')\} &=& \{V(\z), V(\z')\} 
    = \{U(\z), \bar U(\z') \} = 0 \cr
    \{U(\z), V(\z')\} &=& \{V(\z), \bar U(\z')\} = \Ups Z \quad \text{and} \quad
    \{V(\z), -\bar V(\z')\} = 2 (\bar a b \z' - a \bar b \z),
    \label{e:fbp-U-V-anharm-osc-general}
    \eeqs \normalsize
supplemented by their complex conjugates, treating $\z, \z'$ as real (e.g., $\{U(\z), - \bar V(\z')\} = \bar \Ups \bar Z$). Here, we defined $\Ups = 4 \beta h b$.  This can be written compactly as
    \beqs
    \{ A(\z) \stackrel{\otimes}{,} A(\z') \}
    &=& 2 (a \bar b \z' - \bar a b \z) (\sig_- \otimes \sig_+) + 2 (\bar a b \z' - a \bar b \z) (\sig_+ \otimes \sig_-) \cr
    && + \bar \Ups \bar Z ( \sig_3 \otimes \sig_- - \sig_- \otimes \sig_3 ) 
    + \Ups Z ( \sig_3 \otimes \sig_+ - \sig_+ \otimes \sig_3),
    \label{e:fpb-aho-compact-tensor-form}
    \eeqs
where $\sig_\pm = (\sig_1 \pm i \sig_2)/2$ and $\sig_a$ are the Pauli matrices. Note that the second and fourth terms are the adjoints of the first and third, where $(\sig_- \otimes \sig_3)^\dag = \sig_+ \otimes \sig_3$, etc. We wish to express these FPBs in terms of an $r$-matrix, i.e., in the form
    \beq
	\{A(\z) \stackrel{\otimes}{,} A(\z^\pr)\} = [r(\z, \z^\pr), A(\z) \otimes I + I \otimes A(\z^\pr)].
    \eeq
Based on those of generic Zakharov-Shabat systems \cite{babelon-bernard-talon}, we try a nondynamical $r$-matrix of the form
    \beqs
    r(\z, \z^\pr) = \frac{\phi \: P}{(\z - \z^\pr)} \quad \text{where} \quad 
    P(u \otimes v) = v \otimes u
    \label{e:r-matrix-aho-without-rot-egy}
    \eeqs
is the permutation operator and $\phi$ is some function of the parameters. One finds
	\beq
	[r, A(\z) \otimes I + I \otimes A(\z')]_{ijkl} = \frac{\phi}{(\z - \z')} \left(\del_{il}(A_{kj} - A'_{kj}) + \del_{kj}(A'_{il} - A_{il})\right).
    \label{e:comm-r-propto-P-with-A1+A2-index-notation}
	\eeq
Comparing this with the FPBs (\ref{e:fbp-U-V-anharm-osc-general}), we find that they agree iff
    \beq
    \phi = - 4 \beta h b/a = 8 i \mu \beta g \ka_3^3.
    \eeq
% Have checked 1112, 1222, 1222, 1121, 2122 and all other cases and also that $\phi$ reduces to $- \sqrt{2 \beta/\mu}$ in the special case.
The existence of an $r$-matrix implies that traces of powers of $A(\z)$ and $A(\z')$ commute. In fact, proceeding as in (\ref{e:trAp-trAq-pb-with-spectralparam}), and using $[A_1(\z), A_2(\z')] = 0$ we get
    \beq
    \{\tr A^p (\z) \stackrel{\otimes}{,} \tr A^q (\z^\pr) \} 
    = pq \tr A_1(\z)^{p-1} A_2(\z')^{q-1} [r(\z,\z'), A_1(\z) + A_2(\z')] = 0.
    \label{e:trAp-trAq-pb-AHO-with-spectralparam}
    \eeq
Thus, the conserved quantities are in involution.

%-----------------
\section{Fock-Darwin oscillator with quartic potential and Rajeev-Ranken model}
\label{s:iso-FD-osc-quartic-potn-rr-model}
%-----------------

In this section, we extend the results of \S \ref{s:2x2-antiherm-tr-less-lax-pair-iso-anharm-osc} to obtain Lax pairs and associated $r$-matrices for the isotropic Fock-Darwin oscillator with a quartic potential (FD-AHO) and relate them to those of the Rajeev-Ranken (RR) model \cite{rajeev-ranken,gsk-trv-rr-mod-integrability}.

%-----------------
\subsection{Lax pair for isotropic Fock-Darwin oscillator with quartic potential}
\label{s:lax-pair-iso-FD-osc-quartic-potn}
%-----------------

The Lax pair construction of \S \ref{s:2x2-antiherm-tr-less-lax-pair-iso-anharm-osc} may be extended to isotropic anharmonic oscillators with a rotational energy and governed by the Hamiltonian
    \beq
    H_{\fdaho} = \frac{p_x^2 + p_y^2}{2\mu} + \frac{\g (x p_y - y p_x)}{\mu} + \al r^2 + \beta r^4
    \label{e:hamiltonian-fock-darwin-osc-with-quart-potn}
    \eeq
where $r^2 = x^2 + y^2$ and $\g$ is a constant with dimensions of $M/T$. The rotational energy $\g L_z/\mu$ may also be viewed as arising from an electromagnetic field. In fact, Hamilton's equations
	\beqs
	\mu \dot x = p_x - \g y & \text{and} &
	\mu \dot y = p_y + \g x, \cr
	\dot p_x = - \frac{\g}{\mu} p_y - 2 (\al + 2 \beta r^2) x
	& \text{and} &
	\dot p_y = \frac{\g}{\mu} p_x - 2 (\al + 2 \beta r^2) y
	\label{e:hamilton-eq-anh-osc-with-rot-egy}
	\eeqs
written in second order form with $\tl k = 2 \al - \g^2/\mu$,
	\beq
	\mu \ddot x = - (\tl k + 4 \beta r^2) x - 2 \g \dot y \quad
	\text{and} \quad
	\mu \ddot y = - (\tl k + 4 \beta r^2) y + 2 \g \dot x,
	\label{e:aho-with-rot-egy-eom-2nd-order}
	\eeq
are the Newton-Lorentz equations for a particle of mass $\mu$ and charge $q$ moving on the $x$-$y$ plane in a radial electric field and a uniform axial magnetic field:
	\beq
	q \bmE = - (\tlk + 4 \beta r^2) \bmr \quad \text{and} \quad  q \bmB = - 2 \g  \hat z \quad \text{or} \quad q \bmA = - \g \hat \tht = \g (y \hat x - x \hat y)
	\eeq
where $\bmB = \grad \times \bmA$. Although $\g$ contributes to $\bmE$, when $\g = 0$, $\bmB$ is absent. Eq. (\ref{e:hamilton-eq-anh-osc-with-rot-egy}) may also be viewed as those of (i) a Fock-Darwin oscillator \cite{drigho-kuru-negro-nieto-fock-darwin,bandyopadhyay-fock-darwin} subject to a circularly symmetric quartic trapping potential $\beta r^4$ or (ii) the Rajeev-Ranken model restricted to the $x$-$y$ plane (see \S \ref{s:compare-RR-model} and \cite{gsk-trv-rr-mod-anharm-osc}). The presence of four dimensional constants $\mu, \al, \beta$ and $\g$ (with $[\g] = M/T$) permits the construction of one independent dimensionless parameter, say $\g^2/\mu \al$.

In terms of $Z = x + i y$ and $P = p_x + i p_y$, Hamilton's equations (\ref{e:hamilton-eq-anh-osc-with-rot-egy}) become
	\beq
	\mu \dot Z = P + i \g Z \quad
	\text{and} \quad
	\dot P = \frac{i \g P}{\mu} - 2 (\al + 2 \beta |Z|^2)Z \;\;
	\text{or} \quad
	\dot P = i \g \dot Z - 2 (\al + 2 \beta |Z|^2 - \g^2/2\mu) Z.
	\label{e:RR-model-eom-Z-P}
	\eeq

\fl {\bf Ansatz for Lax pair of the FD-AHO.} To find a Lax formulation $\dot A = [B,A]$ for (\ref{e:RR-model-eom-Z-P}) we proceed as in \S \ref{s:2x2-antiherm-tr-less-lax-pair-iso-anharm-osc} by looking for traceless antihermitian $2 \times 2$ matrices $A = (U, V | - \bar V, \bar U)$ and $B = (X, Y | - \bar Y, \bar X)$. However, we need to generalize the ansatz for the matrix entries (\ref{e:V-ansatz-2d-anharm-osc}), (\ref{e:X-ansatz-2d-anharm-osc}) and (\ref{e:Y-U-ansatz-2d-anharm-osc}) to accommodate the $\g$ terms. Introduction of the magnetic field suggests the replacement $P \to P + i \g Z$ (minimal coupling $\bmp - q \bmA$) in $V$ and $Y$ and the replacement $\al \to \al - \g^2/2\mu$ due to its appearance in the electric field via $\tl k$ in (\ref{e:aho-with-rot-egy-eom-2nd-order}) and in (\ref{e:RR-model-eom-Z-P}):
	\beqs
	V &=& a \zeta Z + b (P + i \g Z),  \quad
	X = c \zeta + \frac{d \psi}{\z},
	\cr
	Y &=& \z^{-\del} (e \z Z + f (P + i \g Z))
	\quad \text{and} \quad
	U = \z^{\del} \left( g \z + j + \frac{h}{\z} \psi \right).
	\label{e:UVXY-ansatz-for-lax-RR-model}
	\eeqs
Here $\psi = (\al + 2 \beta |Z|^2 - \g^2/2\mu)$ and $a$ and $b$ are nonzero complex constants. We get
	\beqs
	\dot V &=& a \z \dot Z + b (\dot P + i \g \dot Z), \cr
	VX - UY &=& \z^2(ac - ge)Z + \z((bc-gf)(P + i \g Z) - jeZ) \cr
	&& + ((ad-he)Z \psi - jf(P + i \g Z)) 
	+ \z^{-1} (bd-hf) \psi(P + i \g Z).
	\eeqs

\vspace{4pt}

\fl {\bf Comparison between Hamilton and Lax equations.} For the Lax equation $\dot V = 2 (VX-UY)$ to be equivalent to the EOM  (\ref{e:RR-model-eom-Z-P}),  the coefficients of $\zeta^2$ and $1/\zeta$ must vanish:
	\beq
	ac = ge \quad \text{and} \quad bd = hf.
	\label{e:2-cond-RR-Vdot-lax-eqn-order-zetasq-ovzeta}
	\eeq
The Lax equations at $\calO(\zeta)$ and $\calO(\zeta^0)$ are
	\beq
	a \dot Z = 2(bc - gf) (P + i \g Z) - 2jeZ \quad \text{and} \quad
	b (\dot P + i \g \dot Z) = 2(ad - he) Z \psi - 2 j f (P + i \g Z).
	\eeq
Comparing the $O(\zeta)$ Lax equation with (\ref{e:RR-model-eom-Z-P}) we get the conditions
	\beq
	2 \mu(bc - gf) = a \quad \text{and} \quad j e = 0.
	\label{e:2-cond-RR-Vdot-lax-eqn-order-zeta}
	\eeq
This allows us to express the $O(\zeta^0)$ Lax equation as
	\beq
	\dot P = - \left( \frac{i \g}{\mu} + \frac{2jf}{b} \right) P + Z \left( \frac{2(ad - he)}{b} \psi + \frac{\g^2}{\mu} - \frac{2ijf \g}{b} \right).
	\eeq
Comparing with (\ref{e:RR-model-eom-Z-P}), we get two more conditions:  
	\beq
	i \mu j f = b \g \quad \text{and} \quad
	he - ad = b.
	\label{e:2-cond-RR-Vdot-lax-eqn-order-zeta-0}
	\eeq
Since $b \ne 0$ we must have $j, f \ne 0$ while $e = 0$ and $c = 0$ from (\ref{e:2-cond-RR-Vdot-lax-eqn-order-zeta}) and (\ref{e:2-cond-RR-Vdot-lax-eqn-order-zetasq-ovzeta}).

Turning to the first Lax equation, we find $\dot U = (4 h \beta/\mu) \zeta^{\del - 1} \Re \bar Z P$ as in (\ref{e:U-dot-RR-model}) as $\g$ does not contribute to the radial equation $\mu r \dot r = x p_x + y p_y$. On the other hand, taking $e = 0$,
	\beqs
	V \bar Y &=& \z^{- \del} \bar f [\zeta (a Z \bar P - i a \g |Z|^2) + b(|P|^2 + \g^2 |Z|^2 - 2 \g \Im Z \bar P)] \quad \text{so that}
	\cr
	V \bar Y - \bar V Y &=& 2i \z^{-\del} [\z \{ \Im a \bar f Z \bar P - \g |Z|^2 \Re a \bar f \}
	+ (\Im b \bar f) ( |P|^2 + \g^2 |Z|^2 - 2 \g \Im Z \bar P)	].
	\eeqs
For $\dot U = V \bar Y - \bar V Y$ to be equivalent to the EOM $\mu \dot r = p_r$ the following conditions must be met:
\begin{enumerate}

	\item[(i)] the coefficients of $|P|^2$ and $|Z|^2$ must vanish: $\Im b \bar f = 0$ and  $\Re a \bar f = 0$,

	\item[(ii)] $\del - 1 = - \del + 1$ or $\del = 1$ and
	
	\item[(iii)] $2 h \beta \Re \bar Z P = i \mu \Im a \bar f Z \bar P$.
\end{enumerate}

Condition (i) is equivalent to $f = \ka_2 b$ and $f = i \ka_4 a$  for real constants $\ka_2, \ka_4$. Since $f \ne 0$ we may write $b = i (\ka_4/\ka_2) a$. Putting $a \bar f = -i \ka_4 |a|^2$ in (iii) then gives
	\beq
	2 \beta h = -i \mu \ka_4 |a|^2.
	\eeq
Noting that $c = 0$, the four conditions in (\ref{e:2-cond-RR-Vdot-lax-eqn-order-zeta}) and (\ref{e:2-cond-RR-Vdot-lax-eqn-order-zeta-0}) then give
	\beq
	d = - \frac{i \ka_4}{\ka_2}, \quad
	e = 0, \quad
	g = \frac{i}{2 \mu \ka_4} \quad \text{and} \quad
	j = \frac{\g}{i \mu \ka_2}.
	\eeq
Finally, (\ref{e:2-cond-RR-Vdot-lax-eqn-order-zetasq-ovzeta}) fixes $h$ and the magnitude of $a = |a| e^{i \tht}$:
    \beq
    h = - i \frac{\ka_4}{\ka_2^2} \quad \text{and} \quad 
    |a|^2 = \frac{2 \beta}{\mu \ka_2^2}.
    \eeq
{\it The phase $\tht$ and the nonzero real factors $\ka_2$ and $\ka_4$ may be chosen freely, leading to a $3$-parameter family of such Lax pairs.} In the limit $\g \to 0$, these reduce to a subfamily of the Lax pairs for the quartic anharmonic oscillator of (\ref{e:hamil-quad+quart-anharm-osc}) obtained in \S \ref{s:2x2-antiherm-tr-less-lax-pair-iso-anharm-osc}. In particular, while $e$ must vanish here, it could be nonzero in \S \ref{s:2x2-antiherm-tr-less-lax-pair-iso-anharm-osc}.

\vspace{4pt}

\fl {\bf Sample Lax pair for specific choice of parameters.} A simple choice that extends the one for $\g = 0$ in (\ref{e:coeffs-simple-choice-gamma=0}) is obtained by taking $a$ real, $\ka_4 = 1/8 \beta \mu$ and $\ka_2 = \sqrt{\ka_4}$, giving \small
	\beq
	a = 4 \beta, \;\;
	b = i \sqrt{2 \beta/\mu}, \;\;
	c = 0, \;\;
	d = -i/\sqrt{8 \beta \mu}, \;\;
	e = 0, \;\;
	f = i/2\mu, \;\;
	g = 4 i \beta, \;\;
	h = -i, \;\;
	j = -i \g \sqrt{8 \beta/\mu}.
	\label{e:abcdefghj-coeffs-rr-mod-lax-pair}
	\eeq \normalsize
These correspond to the Lax pair with matrix entries
    \beqs
	U(\z) &=& i \left(4 \beta \z^2 - \g \sqrt{8\beta/\mu} \z - \al - 2 \beta |Z|^2 + \frac{\g^2}{2 \mu} \right), \quad
	V(\z) = 4 \beta \z Z -  \sqrt{2 \beta/\mu} (\g Z - i P)	\cr
	X(\z) &=& -\frac{i}{\z \sqrt{8 \beta \mu}} \left(\al - \frac{\g^2}{2 \mu} + 2 \beta |Z|^2 \right) \quad \text{and} \quad
	Y(\z) = - \ov{2 \z \mu} \left(\g Z - i P \right).
    \label{e:UVXY-corr-to-RR-model-choice}
    \eeqs

%---------------
\subsection{Conserved quantities, FPBs and \texorpdfstring{\boldsymbol{$r$}}{r}-matrix}
\label{s:FD-AHO-FPB-r-matrix-cons-qty}
%---------------

Conserved quantities arise as coefficients of powers of $\z$ in $\tr A^2$. Noting that $\del = 1$, $\Re (i Z \bar P) = L_z$ and that $b \bar a = i (\ka_4/\ka_2)|a|^2$, we find from (\ref{e:UVXY-ansatz-for-lax-RR-model})
	\beqs
	U^2 &=& g^2 \z^4 + 2 g j \z^3 + (j^2 + 2 g h \psi) \z^2 + 2 h j \psi \z + h^2 \psi^2 \quad \text{and} \cr
	\bar V V &=& |a|^2 |Z|^2 \z^2 - 2 |a|^2 (\ka_4/\ka_2) (L_z + \g |Z|^2) \z + |b|^2 (|P|^2 + \g^2 |Z|^2 + 2 \g L_z).
	\eeqs
Using $\psi = \al + 2 \beta |Z|^2 - \g^2/2\mu$ from (\ref{e:UVXY-ansatz-for-lax-RR-model}), we get
    \beqs
    \tr A^2 &=& \frac{2}{\mu^2 \ka_4^2} \bigg[ - \frac{\z^4}{4} + \frac{\g \ka_4}{\ka_2} \z^3 + \frac{\ka_4^2}{\ka_2^2} \left(\mu \al - \frac{3 \g^2}{2} \right) \z^2
    + \frac{\ka_4^3}{\ka_2^3} \left( 4 \beta \mu L_z - 2 \mu \al \g + \g^3 \right) \z
    \cr && - \frac{\ka_4^4}{\ka_2^4} \left( 4 \mu^2 \beta H_{\fdaho} + \mu \al \left(\mu \al - \g^2 \right) + \frac{\g^4}{4} \right) \bigg],
    \eeqs
where $H_{\fdaho} = |P|^2/2\mu + \al |Z|^2 + (\g/\mu) L_z + \beta |Z|^4$ is the Hamiltonian of (\ref{e:hamiltonian-fock-darwin-osc-with-quart-potn}). Up to additive and multiplicative constants, the two conserved quantities $L_z$ and $H_{\fdaho}$ arise as coefficients of $\zeta$ and $\zeta^0$. The coefficients of $\zeta^4, \zeta^3$ and $\zeta^2$ are constants. 

\vspace{4pt}

\fl {\bf Fundamental PBs and \boldsymbol{$r$}-matrix.} The nonzero FPBs between entries of $A(\z) = (U, V | - \bar V, \bar U)$ (\ref{e:UVXY-corr-to-RR-model-choice}) up to complex conjugation are 
    \beqs
    \{ U(\z), V(\z') \} &=& \{ V(\z), \bar U(\z') \} = \Ups Z
    \quad \text{where} \quad \Ups = 4 \beta h b = 4 \beta \sqrt{{2 \beta}/{\mu}} \frac{\ka_4^2}{\ka_2^4} e^{i \tht}.
    \cr
    \{ V(\z), - \bar V(\z') \} &=& 2 (\bar a b \z' - a \bar b \z) - 4 i \g |b|^2
    = \frac{4 i \beta}{\mu} \frac{\ka_4}{\ka_2^3} \left(\z + \z' - 2 \g \frac{\ka_4}{\ka_2} \right).
    \label{e:fbp-FD-U-V-anharm-osc}
    \eeqs
For $\g = 0$, these reduce to the FPBs in (\ref{e:fbp-U-V-anharm-osc-general}). As in (\ref{e:fpb-aho-compact-tensor-form}), they can be written succinctly as
    \beqs
    \{ A(\z) \stackrel{\otimes}{,} A(\z') \}
    &=& ( 2 (a \bar b \z' - \bar a b \z) + 4 i \g |b|^2) (\sig_- \otimes \sig_+) + (2 (\bar a b \z' - a \bar b \z) - 4 i \g |b|^2) (\sig_+ \otimes \sig_-) \cr
    && + \bar \Ups \bar Z ( \sig_3 \otimes \sig_- - \sig_- \otimes \sig_3 ) 
    + \Ups Z ( \sig_3 \otimes \sig_+ - \sig_+ \otimes \sig_3).
    \label{e:fpb-FD-aho-compact-tensor-form}
    \eeqs   

Proceeding as in \S \ref{s:2x2-antiherm-tr-less-lax-pair-iso-anharm-osc}, we express these FPBs in terms of a nondynamical $r$-matrix proportional to the permutation operator:
    \beqs
	\{A(\z) \stackrel{\otimes}{,} A(\z^\pr)\} &=& [r_{\fdaho}(\z, \z^\pr), A(\z) \otimes I + I \otimes A(\z^\pr)] \quad \text{with} \quad 
    r_{\fdaho}(\z, \z^\pr) = \frac{\phi P}{\z - \z'}  \cr
    \text{where} \quad \phi &=& - 4 \beta h b/a = - 4 \beta {\ka_4^2}/{\ka_2^3}.
    \label{e:r-matrix-aho-with-rot-egy}
    \eeqs
As a consequence, by (\ref{e:trAp-trAq-pb-AHO-with-spectralparam}) the traces of powers of $A(\z)$ and $A(\z')$ commute implying that the conserved quantities are in involution.

%------------------
\subsection{Connection to the Rajeev-Ranken model} 
\label{s:compare-RR-model}
%------------------

We now outline an interesting relation between the variables, Hamiltonians, equations, Lax pairs, FPBs and $r$-matrices of the Fock-Darwin anharmonic oscillator (FD-AHO) of (\ref{e:hamilton-eq-anh-osc-with-rot-egy}) and the classical Rajeev-Ranken (RR) model. 

\vspace{4pt}

\fl {\bf The Rajeev-Ranken model.} The RR model describes screw-type waves in a $1+1$ dimensional scalar field theory \cite{rajeev-ranken}, and is governed by six first order ODEs \cite{gsk-trv-rr-mod-integrability} for $L_a(t)$ and $S_a(t)$ with $a = 1,2,3$:
    \beqs
    \dot L_1 = k S_2, & 
    \dot L_2 = - k S_1, &
    \dot L_3 = 0 \quad \text{and} \cr
    \mu \dot S_1 = \la (S_2 L_3 - S_3 L_2), &
    \mu \dot S_2 = \la(S_3 L_1 - S_1 L_3), &
    \mu \dot S_3 = \la(S_1 L_2 - S_2 L_1).
    \label{e:RR-model-eom-SL}
    \eeqs
The wavenumber is related to $k$ while $\la$ is a coupling constant and $\mu$ is a mass. These have been verified to be Hamilton's equations $\dot f = \{ f, H_{\rm rr} \}$ for the step-3 nilpotent PBs
    \beq
    \{ L_a, L_b \} = 0, \quad 
    \{ S_a, S_b \} = (\la/\mu^2) \eps_{abc} L_c, \quad  
    \{ S_a, L_b \} = -(1/\mu) \eps_{abc} K_c
    \label{e:nilpotent-pb-RR-model}
    \eeq
with $K_1 = K_2 = 0, K_3 = -k$ and Hamiltonian 
    \beq
    H_{\rm rr} = \sum\nolimits_{a} (\mu S_a^2 + L_a^2)/2 + \mu k S_3/\la + \mu k^2/2\la^2.
    \label{e:hamil-rr-model}
    \eeq    
Here, $\la$ and $k$ have dimensions of $M^{1/2}/L$ and $M^{1/2}/T$. Although the phase space is six dimensional, this Poisson algebra is degenerate, possessing two independent Casimir invariants that commute with all observables. They can be taken as $L_3$ and ${\cal C} k^2 = \sum_a L_a^2/2\mu + k S_3/\la$. Thus, the dynamics is confined to four-dimensional symplectic leaves which are the common level sets of $L_3$ and $\cal C$. We will imagine using $L_{1,2}$ and $S_{1,2}$ as coordinates on these leaves with $L_3$ and $S_3$ expressed in terms of them using the Casimir invariants. 

\vspace{4pt}

\fl {\bf From \boldsymbol{$L$} and \boldsymbol{$S$} to oscillator variables.} To reformulate the RR model equations as those of an anharmonic oscillator, we write $L_{1,2}$ and $S_{1,2}$ in terms of oscillator variables $x,y,p_x,p_y$:
    \beqs
    L_1 &=& ky, \quad
    L_2 = -kx, \quad
    \mu S_1 = p_x - \g y, \quad
    \mu S_2 = p_y + \g x, \quad \text{and} \cr
    L_3 &=& -mk, \quad
    \mu S_3 = p_z - (\g/m) (x^2 + y^2) - k\mu/\la.
    \label{e:transformation-LS-intermsof-xypxpy}
    \eeqs
The last two equations may be viewed as assigning numerical values ($m, p_z$) to a convenient pair of Casimirs on each symplectic leaf: $-mk = L_3$ and $p_z = \mu \la k({\cal C} - m^2/2\mu + 1/\la^2)$. To make the connection to the FD-AHO (\ref{e:hamiltonian-fock-darwin-osc-with-quart-potn}), we introduce the constants
    \beq
    \al =  \frac{\la^2 k^2 m^2}{8\mu} - \frac{\la k p_z}{2\mu} + \frac{k^2}{2}, \quad
    \beta = \frac{\la^2 k^2}{8\mu} \quad \text{and} \quad 
    \g = \frac{\la m k}{2}.
    \label{e:al-beta-g-la-k-m-relations}
    \eeq
The RR model Hamiltonian (\ref{e:hamil-rr-model}) then becomes
    \beqs
    H_{\rm rr} &=& \frac{\mu}{2} S_a^2 + \mu k^2 \left( {\cal C} + \ov{2 \la^2} \right)
    = \frac{p_x^2 + p_y^2}{2 \mu} + \frac{\g L_z}{\mu} + \frac{\g^2 r^2}{2 \mu} + \frac{\mu}{2} S_3^2 + \mu k^2 \left( {\cal C} + \ov{2 \la^2} \right) \cr
    &=& \frac{p_x^2 + p_y^2}{2 \mu} + \frac{\g L_z}{\mu} + \al r^2 + \beta r^4 + \left[ \frac{p_z^2}{2\mu} + \frac{k^2 m^2}{2} \right] = H_{\fdaho} + \left[ \frac{p_z^2}{2\mu} + \frac{k^2 m^2}{2} \right].
    \eeqs
Up to additive constants, this is the same as the FD-AHO Hamiltonian $H_{\fdaho}$ given in (\ref{e:hamiltonian-fock-darwin-osc-with-quart-potn}). Moreover, it can be written as a sum of squares, 
    \beq
    H_{\rm rr} = \frac{1}{2 \mu} \left [\left(p_x - \frac{\la m k y}{2} \right)^2 + \left( p_y + \frac{\la m k x}{2} \right)^2 + \left( p_z -  \frac{\la k r^2}{2} \right)^2 \right] + \frac{k^2}{2} (r^2 + m^2),
    \label{e:RR-hamiltonian-sum-of-sq}
    \eeq
guaranteeing that $H_{\rm rr} \geq k^2 m^2/2$.

\vspace{4pt}

\fl {\bf Relating RR model and FD-AHO equations.} The RR model equations for $\dot L_1$ and $\dot L_2$ (\ref{e:RR-model-eom-SL}) agree with the $\dot x$ and $\dot y$ equations (\ref{e:hamilton-eq-anh-osc-with-rot-egy}) while the $\dot L_3$ equation is consistent with the constancy of $m$. Using these, the $\dot S_1$ and $\dot S_2$ equations reduce to the $\dot p_x$ and $\dot p_y$ equations while the $\dot S_3$ equation is then identically satisfied. Thus, upon making the change of variables (\ref{e:transformation-LS-intermsof-xypxpy}) from $S$ and $L$ to $x,y,p_x,p_y$, the RR model equations (\ref{e:RR-model-eom-SL}) imply the FD-AHO equations (\ref{e:hamilton-eq-anh-osc-with-rot-egy}).

Inverting (\ref{e:transformation-LS-intermsof-xypxpy}), we may express the oscillator variables $x,y,p_x, p_y$ in terms of $L,S$:
    \beq
    x = - \frac{L_2}{k}, \quad
    y = \frac{L_1}{k}, \quad
    p_x = \mu S_1 + \frac{\g  L_1}{k} \quad
    \text{and} \quad
    p_y = \mu S_2 + \frac{\g L_2}{k},
    \label{e:transformation-xypxpy-in-termsof-L1L2S1S2}
    \eeq
and supplement these by two equations that express the constants $m$ and $p_z$ in terms of the RR model variables $L_3$ and $S_3$:
    \beq
    - m k = L_3 \quad \text{and} \quad
    p_z = \mu S_3 + \frac{\g (x^2 + y^2)}{m} + \frac{k \mu}{\la}.
    \label{e:supplement-m-pz-L3-S3}
    \eeq
The $\dot x$ and $\dot y$ equations (\ref{e:hamilton-eq-anh-osc-with-rot-egy}) imply the $\dot L_2$ and $\dot L_1$ equations (\ref{e:RR-model-eom-SL}) while the $\dot p_x$ and $\dot p_y$ equations supplemented by (\ref{e:supplement-m-pz-L3-S3}) give the $\dot S_1$ and $\dot S_2$ equations. Combining these with the constancy of $p_z$ we get the $\dot S_3$ equation. Thus, the mapping (\ref{e:transformation-xypxpy-in-termsof-L1L2S1S2}) supplemented by (\ref{e:supplement-m-pz-L3-S3}) converts the FD-AHO EOM (\ref{e:hamilton-eq-anh-osc-with-rot-egy}) to the RR model equations (\ref{e:RR-model-eom-SL}). In summary, we may use these transformations to go back and forth between the RR model and the FD-AHO equations of motion. We note in passing, that although $p_z$ arose as a Casimir invariant of the $L$-$S$ Poisson algebra (\ref{e:nilpotent-pb-RR-model}), as the notation suggests, it can also be regarded as a conserved momentum associated to a vertical coordinate $z$ that does not appear in $H_{\rm rr}$ and whose evolution is governed by $\dot z = S_3 + k/\la = p_z/\mu - (\la k/2 \mu)r^2$ (see \S 4.3 of \cite{gsk-trv-rr-mod-integrability} and \S 2 of \cite{gsk-trv-rr-mod-anharm-osc}).

\vspace{4pt}

\fl {\bf Transforming Lax pairs and \boldsymbol{$r$}-matrices from FD-AHO to RR model.} Furthermore, the Lax pair for the FD-AHO obtained in \S \ref{s:lax-pair-iso-FD-osc-quartic-potn} with the choices of Eq. (\ref{e:abcdefghj-coeffs-rr-mod-lax-pair}) turns out to be related to that of the RR model. In fact, upon dividing $A$ (\ref{e:UVXY-corr-to-RR-model-choice}) by $\la^2 k/\mu$ and taking the complex conjugate of $A$ and $B$, we get the RR model Lax pair given in Eq. (50) of \cite{gsk-trv-rr-mod-integrability} in units where $\mu = 1$:
    \beq
    A_{\rm rr}(\z) = ({\mu}/{\la^2 k}) \bar A(\z) = - K \z^2 + L \z + \mu S/\la \quad \text{and} \quad B_{\rm rr}(\z) = \bar B(\z) = {S}/{\z}.
    \label{e:A-B-lax-pair-for-RR-model-original}
    \eeq
Here $L = \ov{2i} L_a \sig_a$, etc., are $2 \times 2$ traceless antihermitian matrices. Thus, our construction has allowed us to find a 3-parameter family of Lax pairs for these circularly symmetric anharmonic oscillators, going beyond the one obtainable from the RR model through a change of variables.

Interestingly, the canonical $x,y,p_x,p_y$ PBs imply the nilpotent $L,S$ PBs of (\ref{e:nilpotent-pb-RR-model}). However, the converse is not quite true. In fact, the PB $\{ S_1, S_2 \} = \la L_3 / \mu^2$ implies $\{p_x, p_y \} = -\la mk$, which is generally nonzero. Incidentally, by picking the Casimir $L_3 = -mk$ to vanish, we recover the canonical PBs between $x, y, p_x, p_y$. Notice that when $mk =0$, $\g =0$ but $\al$ and $\beta $ are nonzero. Notably, although the PBs of the RR model are {\it not} the same as the canonical $x,y,p_x,p_y$ PBs, the RR model FPBs (Eq. (51) of \cite{gsk-trv-rr-mod-integrability}) are related to those of the FD-AHO (\ref{e:fpb-FD-aho-compact-tensor-form}) simply via rescaling by $\la^2 k/\mu$. Thus, the $r$-matrix of the RR model (Eq. (52) of \cite{gsk-trv-rr-mod-integrability}) is a constant multiple of that of the FD-AHO (\ref{e:r-matrix-aho-with-rot-egy}):
    \beq
    r_{\rm rr} (\z, \z') = \frac{\mu}{\la^2 k} r_{\fdaho} = - \frac{P}{2\la(\z - \z')}.
    \label{e:r-mat-rr-model}
    \eeq

\vspace{4pt}

\fl {\bf More general Lax pairs and $r$-matrices for the RR model.} Pleasantly, the 3-parameter family of FD-AHO Lax pairs we found in \S \ref{s:lax-pair-iso-FD-osc-quartic-potn} may be used to construct more general Lax pairs for the RR model. This will allow us to go beyond the $(A,B)$ pair given in (\ref{e:A-B-lax-pair-for-RR-model-original}), which was based on the special choice of parameters in (\ref{e:abcdefghj-coeffs-rr-mod-lax-pair}). To see this, we write the Lax pair entries in (\ref{e:UVXY-ansatz-for-lax-RR-model}) in terms of the components of $L$, $S$ and the free parameters $\ka_2, \ka_4$ and $\tht = \arg a$:
    \beqs
    U &=& i \left( \frac{\z^2}{2 \mu \ka_4} + \frac{\la \z L_3}{2 \mu \ka_2} + \frac{\la k \ka_4 S_3}{2 \ka_2^2} \right), \quad
    V = i e^{i \tht} \left( \frac{\la \z (L_1 + i L_2)}{2 \mu \ka_2}  + \frac{\la k \ka_4 (S_1 + i S_2)}{2 \ka_2^2} \right)
    \cr
    X &=& \frac{i \la k \ka_4}{2 \ka_2 \z} S_3
    \quad \text{and} \quad
    Y = \frac{i \la k \ka_4 e^{i \tht}}{2 \ka_2 \z} (S_1 + i S_2).
    \eeqs
We used $P + i \g Z = \mu(S_1 + i S_2)$ and $\psi = - \la k S_3/2$. The Lax matrices may be written as
    \beqs
    A &=& A_2 \z^2 + A_1 \z + A_0 \quad \text{where}
    \quad A_2 = \frac{i}{2 \mu \ka_4} \sig_3, \quad
    A_1 = \frac{i \la}{2 \mu \ka_2} \begin{smmat} L_3 & e^{i \tht} (L_1 + i L_2) \cr e^{- i \tht} (L_1 - i L_2) & - L_3 \end{smmat}, \cr 
    A_0 &=& \frac{i \la k \ka_4}{2 \ka_2^2} \begin{smmat} S_3 & e^{i \tht} (S_1 + i S_2) \cr e^{- i \tht} (S_1 - i S_2) & -S_3 \end{smmat} \quad 
    \text{and} \quad B = \frac{B_{-1}}{\z} \quad \text{where} \quad
    B_{-1} = \ka_2 A_0.
    \eeqs
To get Lax pairs for the RR model, we divide $A$ (henceforth also referred to as $A_{\fdaho}$) by $\la^2 k/\mu$ and take the complex conjugate of $A$ and $B$:
    \beq
    A_{\rm rr} = \frac{\mu}{\la^2 k} \bar A_{\fdaho}(\z) = \frac{\mu}{\la^2 k} \left( \bar A_2 \z^2 + \bar A_1 \z + \bar A_0 \right) 
    \quad \text{and} \quad 
    B_{\rm rr} = \z^{-1} \bar B_{-1}.
    \label{e:more-gen-Lax-pair-RR-model}
    \eeq
These can be written compactly in a manner analogous to (\ref{e:A-B-lax-pair-for-RR-model-original}):
    \beq
    A_{\rm rr} = - \frac{\z^2}{k^2 \la^2 \ka_4} K + \frac{\z}{k \la \ka_2} L^\tht + \frac{\mu \ka_4}{\la \ka_2^2} S^\tht 
    \quad \text{and} \quad
    B_{\rm rr} = \frac{\la k \ka_4}{\ka_2 \z} S^\tht.
    \label{e:Arr-Brr-gen-params}
    \eeq
Here, $L^\tht$ and $S^\tht$ are obtained from $L$ and $S$ via a rotation by angle $\tht$ in the 1-2 plane, e.g.:
    \beq
    L^\tht = \ov{2i} L^\tht_a \sig_a
    \quad \text{where} \quad L^\tht_a = (L_1 \cos \tht - L_2 \sin \tht, L_1 \sin \tht + L_2 \cos \tht, L_3).
    \eeq
Thus, our new three-parameter family of RR model Lax pairs may be viewed as obtained from the original one (\ref{e:A-B-lax-pair-for-RR-model-original}) by performing a rotation and appropriately rescaling coefficients of various powers of the spectral parameter. Similarly, from (\ref{e:fpb-FD-aho-compact-tensor-form}) we get the FPBs among entries of these new Lax matrices:
\small
    \beqs
    \{ A_{\rm rr}(\z) \stackrel{\otimes}{,} A_{\rm rr}(\z') \}
    &=& \frac{\mu^2}{\la^4 k^2} \{ \bar A_{\fdaho}(\z) \stackrel{\otimes}{,} \bar A_{\fdaho}(\z') \}
    \cr
    &=& \frac{i \ka_4}{2 \la \ka_2^3} \left[ \left( \frac{\z + \z'}{\la} - \frac{m k \ka_4}{\ka_2}  \right) \sig_- \otimes \sig_+ 
    - \frac{\ka_4}{2 \ka_2} L_-^\tht (\sig_3 \otimes \sig_+ - \sig_+ \otimes \sig_3) \right] + \text{h.c.} \quad \quad
    \label{e:fpb-rr-compact-tensor-form}
    \eeqs   \normalsize
where $L_-^\tht = L_1^\tht - i L_2^\tht$. Moreover, we may propose an $r$-matrix for the RR model by transforming that of the FD-AHO from (\ref{e:r-matrix-aho-with-rot-egy}):
    \beq
    r_{\rm rr} = \frac{\mu}{\la^2 k} r_{\fdaho} = \frac{\mu}{\la^2 k} \frac{\phi P}{\z - \z'} = - \frac{k}{2} \frac{\ka_4^2}{\ka_2^3} \frac{P}{\z - \z'}.
    \label{e:r-mat-rr-mod-general-param}
    \eeq
We checked that these Lax pairs, FPBs and $r$-matrices reduce to (\ref{e:A-B-lax-pair-for-RR-model-original}), Eq. (51) of \cite{gsk-trv-rr-mod-integrability} and (\ref{e:r-mat-rr-model}) if we take $\ka_4 = \ka_2^2 = 1/\la^2 k^2$ and $\tht = 0$. More generally, we have verified that the Lax equations that follow from this pair are equivalent to the RR model EOM given in (\ref{e:RR-model-eom-SL}) and that the $r$-matrix  of (\ref{e:r-mat-rr-mod-general-param}) gives the FPBs of (\ref{e:fpb-rr-compact-tensor-form}). Thus, we have found a 3-parameter family of Lax pairs and associated $r$-matrices for the RR model, going beyond the results of \cite{gsk-trv-rr-mod-integrability}.

%--------------------
\section{Conclusion and Discussion}
\label{s:discussion}
%--------------------

In this paper, we have been concerned with Lax pairs $(A,B)$ with spectral parameter $\z$ and $r$-matrices for certain two-dimensional isotropic oscillators. Although the isotropic harmonic oscillator (IHO) can be given a bi-Hamiltonian formulation and also viewed as the limit of a 2-particle harmonic Calogero model, neither of these lead to a Lax pair for it. On the other hand, we find a $4 \times 4$ block diagonal Lax pair for the IHO that gives two of its conserved quantities in involution. Notably, the corresponding block diagonal $r$-matrix is dynamical and depends on only one of the two spectral parameters (in a linear fashion). We also find a family of $2 \times 2$ Lax pairs that lead to all three conserved quantities of the IHO. This is perhaps the simplest example of a Lax pair that gives all conserved quantities of a superintegrable system. They satisfy a nonabelian Poisson algebra, which precludes an $r$-matrix formulation of the fundamental Poisson brackets. In the $4 \times 4$ and $2 \times 2$ constructions, the Lax matrix $A$ is a $\zeta$-weighted direct sum or linear combination of individual oscillator Lax pairs. 

We then construct a family of Lax pairs for the isotropic quartic anharmonic oscillator ($\al r^2 + \beta r^4$) and extend them to the isotropic Fock-Darwin oscillator with a quartic potential ($\al r^2 + \beta r^4 + (\g/\mu) L_z$). In both cases, the fundamental Poisson brackets among Lax matrix entries are expressed in terms of nondynamical rational $r$-matrices. While traceless antihermitian Lax matrices were found to be suitable for the isotropic anharmonic oscillators expressed in complex position and momentum variables, real symmetric-antisymmetric pairs did the job for the linear harmonic oscillator. Interestingly, our anharmonic oscillator Lax pairs are singular in the limit of vanishing anharmonicity. 

Finally, we used a change of variables to transform the anharmonic Fock-Darwin oscillator Lax pairs to obtain a 3-parameter family of Lax pairs for the Rajeev-Ranken model. These are related to a previously derived Lax pair for the RR model via suitable rotations and rescaling of coefficients. Interestingly, the transformed $r$-matrices also work for the RR model since they share the same fundamental PBs despite having distinct Poisson structures.

Although the isotropic oscillators considered in this paper are relatively simple systems, we are not aware of such Lax pairs for them in the existing literature. The methods employed here could also be extended to other interesting systems. What is more, our work raises some interesting technical and conceptual questions. {\bf (1)} If the conserved quantities arising from a Lax matrix satisfy a nonabelian Poisson algebra, can this be encoded in an expression for the fundamental Poisson brackets among Lax matrix elements, just as an $r$-matrix makes this possible when the algebra is abelian? {\bf (2)} Is there a Lax pair for the Fock-Darwin linear oscillator without any quartic potential? {\bf (3)} Can we find Lax pairs for the quartic isotropic anharmonic oscillators of (\ref{e:hamil-quad+quart-anharm-osc}) and (\ref{e:hamiltonian-fock-darwin-osc-with-quart-potn}) that in the limit of vanishing anharmonicity lead to Lax pairs for the isotropic harmonic oscillator and Fock-Darwin oscillators? If not, is there some obstruction to finding such Lax pairs? Our Lax pairs are singular in the harmonic limit. {\bf (4)} It would be interesting to use our anharmonic oscillator Lax pairs and $r$-matrices to formulate separation of variables \cite{eilbeck-enolskij-kuznetsov,dubrovin-skrypnyk-sov-r-matrix}. {\bf (5)} Can one formulate the superintegrability of a system using a Lax matrix whose spectral invariants give all conserved quantities? {\bf (6)} Although we have presented Lax pairs for 2d quartic isotropic anharmonic oscillators, we have not considered the corresponding 1d quartic AHO, for which we are not aware of a Lax pair. {\bf (7)} It would be interesting to examine the analogue of the classical Yang-Baxter equation for the dynamical $4 \times 4$ $r$-matrix we have proposed for the isotropic harmonic oscillator. {\bf (8)} It would also be interesting to understand the dynamics of the models in terms of the spectral curves associated to the Lax pairs we have found. 

\vspace{5pt}

\fl {\bf Acknowledgements:} We would like to thank T R Vishnu for his helpful comments on the manuscript. This work was supported in part by the Infosys Foundation.

\appendix

%-----------------------
\section{FPBs and \texorpdfstring{\boldsymbol{$r$}}{r}-matrix for \texorpdfstring{\boldsymbol{$4 \times 4$}}{4 x 4} isotropic harmonic oscillator Lax matrix }
\label{a:fpb-r-mat-for-4x4-isotropic-sho-lax}
%-----------------------

\paragraph{}In \S \ref{s:4x4-block-lax-pair-iso-harm-osc} we found a $4 \times 4$ block-form Lax matrix for the isotropic 2d harmonic oscillator. The fundamental Poisson brackets between Lax matrix entries was given in (\ref{e:fpb-spec-para-2d-sho}). To find an $r$-matrix formulation of these FPBs, we will use a direct sum decomposition of the auxiliary vector spaces $V$ and $W = V \otimes V$ on which $A$ and the $r$-matrix act. 

\paragraph{Tensor product notation.} Let $V_\pm$ denote the real 2d vector spaces on which $A_\pm$ and $B_\pm$ act. Then $A = A_+ \oplus \z A_-$ and $B = B_+ \oplus B_-$ act on the $n = 4$ dimensional vector space $V = V_+ \oplus V_-$. Let $E_{ij}$ for $i,j = 1,\ldots, n$ denote the canonical $n^2 = 16$ dimensional basis for linear operators on $V$, with matrix elements $(E_{ij})_{kl} = \del_{ik} \del_{jl}$. Then $A = \sum_{ij} A_{ij} E_{ij}$. A basis for linear operators on $V \otimes V$ is given by $E_{ij} \otimes E_{kl}$. Given a pair of operators $C, D: V \to V$ the components of their tensor product in this basis are denoted $(C \otimes D)_{ij,kl} =  C_{ij} D_{kl}$, where $j,l$ are domain indices and $i,k$ are target indices. For example, the two slot-wise embeddings of $A$ as linear operators on $W$: $A_1 = A \otimes I$ and $A_2 = I \otimes A$, have the entries $(A_1)_{ij,kl} = A_{ij} \del_{kl}$ and $(A_2)_{ij,kl} = \del_{ij} A_{kl}$. The trace is given by $\tr C \otimes D = C_{ii} D_{kk}$. The components of the composition of a pair of operators $R,S : W  \to W$ are $(RS)_{ij,kl} = R_{im,kn} S_{mj,nl}$. $C \otimes D$ may also be represented as an $n \times n$ block matrix where each block is an $n \times n$ matrix. In the expression $(A \otimes B)_{ij,kl} =  A_{ij} B_{kl}$, the indices $i,j$ label the blocks and $k,l$ are the indices that denote rows and columns within a block. For $n = 2$, we have $A \otimes B = (A_{11} B, A_{12} B | A_{21} B, A_{22} B)$.

\paragraph{Direct sum decomposition of \boldsymbol{$W$} and entries of \boldsymbol{$A_1$}, \boldsymbol{$A_2$} and \boldsymbol{$r_{12}$}.}For the 2d oscillator, we may decompose $W = V \otimes V$ as the direct sum of four subspaces
	\beq
	W = (V_+ \otimes V_+) \oplus (V_- \otimes V_-) \oplus (V_+ \otimes V_-) \oplus (V_- \otimes V_+).
	\label{e:dir-sum-decomp-VxV-4-subsp}
	\eeq
Noting that the identity on $V_+ \oplus V_-$ may be written as $I_+ \oplus I_-$, we observe that $A_1$ and $A_2$ are block diagonal with respect to this decomposition:
    \beqs
    A_1(\z) &=& A \otimes I = \diag(A_+ \otimes I_+, \z A_- \otimes I_-, A_+ \otimes I_-, \z A_- \otimes I_+) 
    \quad \text{and} \cr
    A_2(\z) &=& I \otimes A = \diag(I_+ \otimes A_+, \z I_- \otimes A_-, \z I_+ \otimes A_-, I_- \otimes A_+).
    \eeqs
The $r$-matrices $r_{12}$ and $r_{21}$, being linear operators on $W$, can be written as $4 \times 4$ block matrices, where each entry is a linear operator between subspaces in the decomposition (\ref{e:dir-sum-decomp-VxV-4-subsp}), ordered as $++, --, +-, -+$. When its entries are written as $r_{ij,kl}$, index values $1,2$ will label a basis for $V_+$ while $3,4$ refer to $V_-$. For example, $r_{23,14}$ lies in the $+-,+-$ block and goes form the $--$ to $++$ subspace of $W$. The desired $r$-matrix formulation of the FPBs (\ref{e:fpb-spec-para-2d-sho}) is
	\beq
	\{ A_1(\z), A_2(\z') \} = [r_{12}(\z, \z'), A_1(\z)] - [r_{21}(\z, \z'), A_2(\z')].
	\label{e:fpb-rmatrix-spec-para-appendix}
	\eeq
where antisymmetry is guaranteed if
    \beq
	P r_{12}(\z, \z') P^{-1} = r_{21}(\z', \z)
    \quad 
    \text{or} \quad
    r_{12}(\z, \z')_{ij,kl} = r_{21}(\z',\z)_{kl,ij} \quad 
    \text{where} \quad 
    P_{ij,kl} = \del_{il} \del_{kj}. 
	\label{e:r21-r12-with-spec-param}
	\eeq
Alternatively, we could replace (\ref{e:r21-r12-with-spec-param}) with $P r_{21}(\z, \z') P^{-1} = r_{12}(\z, \z')$ but reverse the order of spectral parameters in the 2nd term of (\ref{e:fpb-rmatrix-spec-para-appendix}), i.e., use $r_{21}(\z', \z)$.

Inspired by (\ref{e:r-matrix-1d-sho}), we propose an $r$-matrix whose nonzero blocks act on the $++$ and $--$ subspaces to ensure that entries of $A_+$ and $A_-$ Poisson commute:
	\beqs
	r_{12}(\z, \z') &=& \diag(\ka_+ S \otimes A_+, \ka_-f(\z,\z') S \otimes A_-,0,0) \cr
	\text{and} \quad
	r_{21}(\z, \z') &=& \diag(\ka_+ A_+ \otimes S, \ka_-g(\z,\z') A_- \otimes S, 0, 0)
    \quad \text{where} \quad S = \begin{smmat} 0 & 1 \\ -1 & 0 \end{smmat}.
	\label{e:r12-r21-with-spec-para}
	\eeqs
Here, $\ka_\pm = -\om/4 H_\pm$ while $f$ and $g$ are suitable functions. Imposing (\ref{e:r21-r12-with-spec-param}), we find that $f(\z', \z) = g(\z, \z')$. Substituting, the RHS of (\ref{e:fpb-rmatrix-spec-para-appendix}) becomes
    \beq
    \diag(\ka_+(C_+ \otimes A_+ - A_+ \otimes C_+), \ka_-(\z f(\z,\z')C_- \otimes A_- - \z' f(\z',\z) A_- \otimes C_-), 0, 0),
    \label{e:commutator-spec-para}
    \eeq
where $C_\pm = [S, A_\pm] = 2 \begin{smmat} \om x_\pm & - p_\pm/\mu \cr - p_\pm/\mu &  -\om x_\pm \end{smmat}$. The $(++,++)$ block of (\ref{e:commutator-spec-para}) inserted in (\ref{e:fpb-rmatrix-spec-para-appendix}) gives the first two FPBs in (\ref{e:fpb-spec-para-2d-sho}). Comparing the $(--,--)$ block with the $\{A_{34}(\z), A_{33}(\z')\}$ FPB, we get the following condition on $f$:
	\beqs
	\z \z' \om/\mu 
%    &=& ([r_{12}(\z,\z'), A_1(\z)] - [ r_{21}(\z,\z'), A_2(\z')])_{34,33} \cr
	&=& \ka_-(\z f(\z,\z')C_{-34} A_{-33} - \z' f(\z',\z)A_{-34} C_{-33})
	\cr
	&=& (\om/2 \mu^2 H_-) \: \left[ \z f(\z,\z') p_-^2 + \mu^2 \om^2 \z' f(\z',\z) x_-^2 \right].
	\eeqs
This can be satisfied if we pick $f(\z,\z') = \z'$. We checked that this choice also gives us the remaining FPBs in (\ref{e:fpb-spec-para-2d-sho}). 
%For example, \small
%	\beqs
%	([r_{12}(\z,\z'), A_1(\z)] - [ r_{21}(\z,\z'), A_2(\z')])_{43,33} 
%	&=&  \ka_-(\z f(\z,\z')C_{-43} A_{-33} - \z' f(\z',\z)A_{-43} C_{-33})\cr
%	&=& - \ka_- (2 \z \z') (p_-^2/\mu^2 + \om^2 x_-^2) = \z \z' \om/\mu,
%	\eeqs \normalsize
%which agrees with (\ref{e:fpb-spec-para-2d-sho}). 
Thus, we arrive at a block diagonal $r$-matrix for the Lax pair in (\ref{e:4x4-lax-pair-iso-harm-osc}):
    \beqs
    r_{12}(\z,\z') &=& \diag(\ka_+ S \otimes A_+, \: \z' \ka_- S \otimes A_-, \: 0, \: 0) \quad \text{and} \cr
    r_{21}(\z,\z') &=& \diag(\ka_+ A_+ \otimes S, \; \z \ka_- A_- \otimes S, \: 0, \: 0).
    \eeqs
This is discussed further in \S \ref{s:4x4-fpb-r-matrix}.

%-----------------------


\begin{thebibliography}{99}
%-----------------------

\footnotesize

\bibitem{fock-1928} V. Fock, {\it Bemerkung zur Quantelung des harmonischen Oszillators im Magnetfeld}, Z. Phys. {\bf 47}, 446 (1928). 

\bibitem{darwin-1930} C. G. Darwin, {\it The diamagnetism of the free electron}, Proc. Cambridge Philos. Soc. {\bf 27}, 86 (1930). 

\bibitem{rajeev-ranken} S. G. Rajeev and E. Ranken, {\it Highly nonlinear wave solutions in a dual to the chiral model}, Phys. Rev. D $\mathbf{93}$, 105016 (2016).

\bibitem{gsk-trv-rr-mod-anharm-osc} G. S. Krishnaswami and T. R. Vishnu, {\it Quantum Rajeev-Ranken model as an anharmonic oscillator}, J. Math. Phys. {\bf 63}, 032101 (2022).

\bibitem{quantum-dot-anharm-osc} M. Luban, J. H. Luscombe, M. A. Reed and D. L. Pursey, {\it Anharmonic oscillator model of a quantum dot nanostructure}, App. Phys. Lett. {\bf 54}, 1997 (1989).

\bibitem{2d-harm-trap-bose-gas} B. P. van Zyl, R. K. Bhaduri and J. Sigetich, {\it Dilute Bose gas in a quasi-two-dimensional trap}, J. Phys. B: At. Mol. Opt. Phys., {\bf 35}, 1251 (2002).

\bibitem{drigho-kuru-negro-nieto-fock-darwin} E. Drigho-Filho, S. Kuru, J. Negro and L. M. Nieto, {\it Superintegrability of the Fock-Darwin system}, Ann. Phys. (N.Y.) {\bf 383}, 101 (2017). 

\bibitem{sinha-ghosh-bagchi-anisotropic-osc-2023} A. Sinha, A. Ghosh and B. Bagchi, {\it Dynamical symmetries of the anisotropic oscillator}, Phys. Scr. 98, 095253 (2023), \href{https://arxiv.org/abs/2304.14306}{arXiv:2304.14306}.

\bibitem{peter-lax} P. Lax, {\it Integrals of nonlinear equations of evolution and solitary waves}, Comm. Pure Appl. Math. {\bf 21}, 467 (1968).

\bibitem{flaschka} H. Flaschka, {\it The Toda lattice. II. Existence of integrals}, Phys. Rev. {\bf B}9, 1924 (1974).

\bibitem{calogero} F. Calogero, {\it Solution of One-Dimensional N-Body Problems with Quadratic and/or Inversely Quadratic Pair Potentials}, J. Math. Phys. {\bf 12}, 419 (1971).

\bibitem{sutherland-1971} B. Sutherland, {\it Exact Results for a Quantum Many-Body Problem in One Dimension}, Phys. Rev. A {\bf 4}, 2019 (1971).

\bibitem{sutherland-1972} B. Sutherland, {\it Exact Results for Quantum Many-Body Problem in One Dimension. II.}, Phys. Rev. A {\bf 5}, 1372 (1972).

\bibitem{moser} J. Moser, {\it Three integrable Hamiltonian systems connnected with isospectral deformations}, Adv. Math. {\bf 16}, 197 (1975).

\bibitem{arutyunov-book} G. Arutyunov, {\it Elements of Classical and Quantum Integrable Systems}, Springer Nature Switzerland AG (2019).

\bibitem{hoppe-book} J. Hoppe, {\it Lectures on Integrable Systems}, Springer-Verlag, Berlin (1992).

\bibitem{avan-talon-neumann-model} J. Avan and M. Talon, {\it Poisson structures and integrability of the Neumann-Moser-Uhlenbeck model}, Int. J. Mod. Phys. {\bf A}5, 4477 (1989).

\bibitem{kudryashov} N. A. Kudryashov, {\it Lax pair and first integrals for two of nonlinear coupled oscillators}, Advanced Technologies in Robotics and Intelligent Systems, Proceedings of ITR {\bf 80} (2019), Springer, Cham (2020), % \href{https://arxiv.org/pdf/2101.12042}{https://arxiv.org/pdf/2101.12042}.

\bibitem{babelon-bernard-talon} O. Babelon, D. Bernard and M. Talon, {\it Introduction to Classical Integrable Systems}, Cambridge University Press, New York (2003).

\bibitem{gsk-trv-rr-mod-integrability} G. S. Krishnaswami and T. R. Vishnu, {\it On the Hamiltonian formulation, integrability and algebraic structures of the Rajeev-Ranken model}, J. Phys. Commun. {\bf 3}, 025005 (2019), \href{https://arxiv.org/abs/1804.02859}{arXiv:1804.02859}.

\bibitem{eilbeck-enolskij-kuznetsov} J. Eilbeck, V. Z. Enol’skij, V. B. Kuznetsov, A. V. Tsiganov, {\it Linear $r$-matrix algebra for classical separable systems}, J. Phys. A, 27, 567 (1994). % arXiv:hep-th/9306155

\bibitem{dubrovin-skrypnyk-sov-r-matrix} B. Dubrovin and T. Skrypnyk, {\it Separation of variables for linear Lax algebras and classical $r$-matrices}, J. Math. Phys. {\bf 59}, 091405 (2018).

\bibitem{das-integ-models} A. Das, {\it Integrable Models}, World Scientific (1989).

\bibitem{ral-fernandes} R. A. L. Fernandes, {\it Completely integrable bi-Hamiltonian systems}, PhD Thesis, Univ of Illinois Urbana-Champaign (1994).
% http://publish.illinois.edu/ruiloja/files/2023/07/Thesis.pdf

\bibitem{magri} F. Magri, {\it A simple model of the integrable Hamiltonian equation}, J. Math. Phys. {\bf 19} 1156 (1978).

\bibitem{magri-morosi} F. Magri and C. Morosi, {\it A geometric characterization of the integrable Hamiltonian systems through the theory of Poisson Nijenhuis manifolds}, Publ. Universita degli
Studi de Milano (1984).

\bibitem{abanov-gromov-kulkarni} A. Abanov, A. Gromov and M. Kulkarni, {\it Soliton solutions of a Calogero model in a harmonic potential}, J. Phys. A. {\bf 44}, 21 (2011). % https://arxiv.org/abs/1103.6231

\bibitem{babelon-viallet-1990} O. Babelon and C-M. Viallet, {\it Hamiltonian structures and Lax equations}, Phys. Lett. B 237, 411 (1990). % DOI: 10.1016/0370-2693(90)91198-K

\bibitem{bandyopadhyay-fock-darwin} M. Bandyopadhyay, {\it Orbital magnetism of two-dimensional electron gas in a crossed electromagnetic field: the effect of spin-orbit interaction, confined geometries and defects}, J. Stat. Mech. 2006, P10010 (2006). 

\end{thebibliography}
\end{document}